\documentclass{ws-mpla}
\usepackage{url}
\usepackage{subfigure}
\usepackage{graphicx,color,dcolumn}
\usepackage{url}
\RequirePackage{color}

\begin{document}

\markboth{Authors' Names}
{Instructions for Typing Manuscripts (Paper's Title)}

\catchline{}{}{}{}{}

\title{PRESSURE OF DEGENERATE AND RELATIVISTIC ELECTRONS IN  A SUPERHIGH MAGNETIC FIELD }

\author{\footnotesize ZHI FU, GAO}

\address{1. Xinjiang Astronomical Observatory, CAS, 150, Science
1-Street, Urumqi, Xinjiang,\\ 830011, China. zhifu$_{-}$gao@xao.ac.cn\\
2. Key Laboratory of Radio Astronomy, CAS, Nanjing, Jiangshu, 210008, China }

\author{NA, WANG}
\address{Xinjiang Astronomical Observatory, CAS, 150, Science 1-Street,Urumqi Xinjiang,830011, China}
\author{ QIU HE, PENG}
\address{Shchool of Astronomy and Space Science, Nanjing University, 22, Hankou\\ Road, Nanjing, Jiangshu, 210093, China}
\author{XIANG DONG, LI}
\address{Shchool of Astronomy and Space Science, Nanjing University, 22, Hankou \\Road, Nanjing, Jiangshu, 210093, China}
\author{YUAN JIE, DU}
\address{National Space Science Center, CAS, 1-Nanertiao, Zhongguancun, \\Haidian District, Beijing, 100190, China}
\maketitle

\pub{Received (Day Month Year)}{Revised (Day Month Year)}

\begin{abstract}
Based on our previous work, we deduce a general formula for pressure of
degenerate and relativistic electrons, $P_{e}$, which is suitable for
superhigh magnetic fields, discuss the quantization of Landau levels of
electrons, and consider the quantum electrodynamic(QED) effects on the
equations of states (EOSs) for different matter systems. The main
conclusions are as follows: $P_{e}$ is related to the magnetic field $B$,
matter density $\rho$, and electron fraction $Y_{e}$; the stronger the
magnetic field, the higher the electron pressure becomes; the high electron
pressure could be caused by high Fermi energy of electrons in a superhigh
magnetic field; compared with a common radio pulsar, a magnetar could be
a more compact oblate spheroid-like deformed neutron star due to the
anisotropic total pressure; and an increase in the maximum mass of a
magnetar is expected because of the positive contribution of the magnetic
field energy to the EOS of the star.
\end{abstract}

\ccode{PACS: 71.70.Di; 97.0.L.d;  71.18.+y.}

\section{Introduction}	
Thompson and Duncan (1996) predicted that superhigh magnetic fields could
exist in the interiors of magnetars, which are powered by extremely strong
magnetic fields, $B\sim 10^{14}$ to $10^{15}$~G (see Ref.~\refcite{Thompson96}).
The majority of magnetars are classified into two populations historically:
the soft gamma-ray repeaters (SGRs), and the anomalous X-ray pulsars (AXPs).
Pulsars have been recognized to be normal neutron stars (NSs), but sometimes
have been argued to be quark stars (e.g., see Ref.~\refcite{Xu02},~\refcite{Xu05},
~\refcite{Du09}).

It is universally accepted that ${}^3P_2$ anisotropic neutron superfluid could
exist in the interior of a neutron star (NS). If the superhigh magnetic fields of magnetars
originate from the magnetic fields induced by the ferromagnetic moments of the
${}^3P_2$ Cooper pairs of the anisotropic neutron superfluid at a moderate to
lower temperatures ($T \ll 2.87\times 10^{8}$ K, the critical temperature of
the ${}^3P_2$ neutron superfluid) and high nuclear density ($\sim 0.5 \rho_{0}
< \rho < 2.0 \rho_{0}$), then the maximum magnetic field strength for the heaviest
magnetar may be estimated to be $(3.0\sim 4.0)\times 10^{15}$ G, according to our
model (see Ref.~\refcite{Peng07},~\refcite{Peng09}).

For completely degenerate ($T\rightarrow 0$, i.e., $\mu/kT \rightarrow \infty$,
$\mu$ is the chemical potential of species $i$, also called the Fermi energy,
$E_{\rm F}(i)$) and relativistic electrons in equilibrium, the distribution function
$f(E_{e})$ can be expressed as
\begin{equation}
   f(E_{e})= \frac{1}{Exp[(E_{e}- \mu_{e})/kT]+ 1}~~,~~
   \label{1}
\end{equation}

where the sign $+$ refers to Fermi-Dirac statistics, $k$ represents Boltzmann's
constant; and $\mu_{e}$ is the electron chemical potential; when $E_{e}
\leq E_{\rm F}(e), f(E_{e})= 1$; when $E_{e}> E_{\rm F}(e)$,
$f(E_{e})= 0$. The electron Fermi energy $E_{\rm F}(e)$ has the simple
form
\begin{equation}
E_{\rm F}^{2}(e) = p^{2}_{\rm F}(e)c^{2} + m^{2}_{e}c^{4}~~,~~
\label{2}
\end{equation}
with $p_{\rm F}(e)$ being the electron Fermi momentum.

In the interior of a NS, when $B$ is too weak to be taken into consideration,
$p_{\rm F}(e)$, is mainly determined by matter density $\rho$ and the
electron fraction $Y_{e}$ (see Ref.~\refcite{Shapiro83}). In the weak-field
limit $B^{*}\ll 1$ (${B^{*}= B/B_{\rm cr}}$ and $B_{\rm cr}= 4.414 \times 10^{13}$ G
is the electron critical field), the equation of state (EOS) can be written in
the polytropic form,
 \begin{equation}
 P_{e} = K \rho^{\Gamma}~~,~~
 \label{3}
\end{equation}
where $K$ and $\Gamma$ are constants, in the following two limiting cases:
1.For non-relativistic electrons, $\rho\ll 10^{6}$~g~cm$^{-3}$,
\begin{equation}
     \Gamma= \frac{5}{3}, ~~K = \frac{1.0036\times10^{13}}{(A/Z)^{5/3}} ~~~ {\rm cgs}~~,~~
      \label{4}
    \end{equation}
where $A$ and $Z$ are the number of nucleons and the number of protons, respectively. 				
2. For extremely relativistic electrons, $\rho \gg 10^{6}$~g~cm$^{-3}$,
 \begin{equation}
     \Gamma= \frac{4}{3},~~~ K = \frac{1.2435\times 10^{15}}{(A/Z)^{4/3}}  ~~~~~{\rm cgs}~.~~
      \label{5}
    \end{equation}	 					 		
Be note that, for a given nucleus with proton number $Z$ and nucleon number $A$, the
relation of $Y_{e}= Y_{p}= Z/A$ always approximately holds, where $Y_{p}$ is the proton
fraction; for an ideal neutron-proton-electron ($npe$) gas, $Y_{e} = Y_{p}=
\frac{n_{e}}{n_{B}}=\frac{n_{e}}{n_{p} + n_{n}}\approx \frac{n_{e}}
{n_{n}}$, where $n_{e}$, $n_{p}$, $n_{n}$ and $n_{B}$ are the electron
number density, proton number density, neutron number density and baryon number density,
respectively.

In this paper, we focus on the interior of a magnetar where electrons are degenerate
and relativistic. The effects of a strong magnetic field on the equilibrium
composition of a NS have been shown in detail in previous studies (e.g., Ref.~\refcite
{Yakovlev01},~\refcite{Lai91}). In accordance with the popular point of view on the
electron pressure in strong magnetic fields, the stronger the magnetic field, the lower
the electron pressure becomes.  With respect to this viewpoint, we cannot directly
verify it by the experiment in actual existence, owing to the lack of such high-value
magnetic fields on the earth. After a careful check, we found that popular methods of
calculating the Fermi energy of electrons are contradictory to the quantization of
electron Landau levels. In an extremely strong magnetic field, the Landau column
becomes a very long and very narrow cylinder along the magnetic field. By introducing
the Dirac $\delta$-function, we obtain
\begin{equation}
\frac{4\pi}{3B^{*}}(\frac{m_{e}c}{h})^{3}(\gamma_{e})^{4}\int_{0}^{1}
(1- \frac{1}{\gamma^{2}_{e}}- \chi ^{2})^{\frac{3}{2}}d\chi
-2\pi\gamma_{e}(\frac{m_{e}c}{h})^{3}\sqrt{2B{*}}= N_{A}\rho Y_{e}~~,~~
\label{6}
\end{equation}
where $\chi$ and $\gamma_{e}$ are two non-dimensional variables,
defined as $\chi=(\frac{p_{z}}{m_{e}c})/(\frac{E_{\rm F}(e)}{m_{e}
c^{2}})= p_{z}c/E_{\rm F}(e)$ and $\gamma_{e}= E_{\rm F}(e)/m_{e}c^{2}$,
respectively; $1/\gamma^{2}_{e}$ is the modification factor; $h$ and $N_{A}$
are the Planck constant and Avogadro constant, respectively (see Ref.~
\refcite{Gao11a},~\refcite{Gao12a}). Solving Eq.(6) gives a concise formula
for $E_{\rm F}(e)$ in superhigh magnetic fields,
\begin{equation}
E_{\rm F}(e)\simeq 43.44(\frac{B}{B_{\rm cr}})^{1/4}(\frac{\rho}{\rho_{0}}\frac{Y_e}{0.0535})^{\frac{1}{4}}~~~~
~\rm MeV~~,~\label{7}
\end{equation}
where $\rho_{0}= 2.8\times 10^{14}$~g~cm$^{3}$ is the standard nuclear
density (see Ref.~\refcite{Gao12a}).

The remainder of this paper is organized as follows: in Section 2, on the basis of our
previous work, we deduce an equation involving $P_{e}$, $\rho$, $B$ and $Y_{\rm e}$,
which is suitable for strong magnetic fields; in Section 3, we take into account QED effects
on EOSs of different matter systems, and discuss an anisotropy of the total pressure of
ideal $npe$ gas due to strong magnetic fields; in Section 4, we present a dispute on
$P_{e}$ in superhigh magnetic fields, and finally we summarize our findings
with conclusions in Section 5.
\section{Pressure of degenerate and relativistic electrons}
 The relativistic Dirac-Equation for the electrons in a uniform
external magnetic field along the $z-$axis gives the electron energy level
\begin{equation}
E_{e}= [m_{e}^{2}c^{4}(1+ \nu \frac{2B}{B_{\rm cr}})+ p^{2}
_{z}c^{2}]^{\frac{1}{2}}~~, ~~\label{8}
\end{equation}
where the quantum number $\nu$ is given by $\nu= n + \frac{1}{2}+
\sigma$ for the Landau level $n= 0, 1, 2, \cdots $, spin $\sigma
=\pm \frac{1}{2}$ (see Ref.~\refcite{Canuto77}), and the quantity $p_{z}$ is the
$z$-component of the electron momentum and may be treated as a
continuous function.  Combining $B_{\rm cr}= m^{2}_{e}c^{3}/{e}\hbar$
with $\mu_{e}^{'} = e\hbar/2m_{e}c$ gives
\begin{equation}
  E_{e}^{2}= m_{e}^{2}c^{4} + p_{z}^{2}c^{2}
 + 2\nu 2m_{e}c^{2}\mu_{e}^{'}B~~, ~~\label{9}
  \end{equation}
where $\mu_{e}^{'}$ is the magnetic moment of an electron.

For the convenience of the following calculations, we define the electron
momentum perpendicular to the magnetic field, $p_{\bot}=m_{e}c(2\nu B^{*})
^{\frac{1}{2}})$.  Then, Eq.(9) can be rewritten as,
\begin{equation}
  E_{e}^{2}= m_{e}^{2}c^{4} + p_{z}^{2}c^{2}
 + p_{\bot}^{2}c^{2}~~, ~~\label{10}
  \end{equation}
The maximum electron Landau level number $n_{max}$ is uniquely determined
by the condition $[p_{\rm F}(z)c]^{2}\geq 0$ (see Ref.~\refcite{Lai91}), where
$p_{\rm F}(z)$ is the Fermi momentum along the $z-$axis.  The
expression for $n_{max}$ is
\begin{eqnarray}
 && n_{\rm max}(\sigma= -\frac{1}{2}) = Int[\frac{1}{2B^{*}}[(\frac{E_{\rm F}(e)}{m_{e}c^{2}})
 ^{2} -1 -(\frac{p_{z}}{m_{e}c})^{2}]]~~,\nonumber\\
 && n_{\rm max}(\sigma=~\frac{1}{2}) = Int[\frac{1}{2B^{*}}[(\frac{E_{\rm F}(e)}{m_{e}c^{2}})
 ^{2} -1 -(\frac{p_{z}}{m_{e}c})^{2}]-1]~~,
 ~~\label{11}
\end{eqnarray}

where $Int[x]$ denotes an integer value of the argument $x$. Correspondingly, the
expression for $\nu_{max}$ can be expressed as
\begin{eqnarray}
 && \nu_{max}(\sigma=-\frac{1}{2})= Int[\frac{1}{2B^{*}}[(\frac{E_{\rm F}(e)}{m_{e}c^{2}})
 ^{2}-1 -(\frac{p_{z}}{m_{e}c})^{2}]+\frac{1}{2}-\frac{1}{2}]~~,\nonumber  \\
 &&= Int[\frac{1}{2B^{*}}[(\frac{E_{\rm F}(e)}{m_{e}c^{2}})
 ^{2}-1-(\frac{p_{z}}{m_{e}c})^{2}]]~~,\nonumber\\
&& \nu_{max}(\sigma=\frac{1}{2})= Int[\frac{1}{2B^{*}}[(\frac{E_{\rm F}(e)}{m_{e}c^{2}})
 ^{2}-1-(\frac{p_{z}}{m_{\rm e}c})^{2}]-1+\frac{1}{2}+~\frac{1}{2}]~~,\nonumber\\
 &&=Int[\frac{1}{2B^{*}}[(\frac{E_{\rm F}(e)}{m_{e}c^{2}})
 ^{2}-1-(\frac{p_{z}}{m_{e}c})^{2}]]~~.~~\label{12}
\end{eqnarray}
From Eq.(12), it's obvious that
\begin{equation}
\nu_{max}(\sigma=-\frac{1}{2})=\nu_{max}(\sigma= \frac{1}{2}).
\label{13}
\end{equation}

However, the electron Fermi momentum is the maximum of electron momentum.
The physics on the condition $[p_{\rm F}(z)c]^{2}\geq 0$ in Ref.~\refcite{Lai91}
includes that: $n_{max}$ is determined by $E_{\rm F}(e)$ and $p_{z}$
when $B$ is given; for a given Landau level number $n$, $p_{z}$ always has the
maximum $p_{z}(max)$ corresponding to $n$; in the same way, when $p_{z}$ is
given, $n$ also has the maximum $n_{max}$ corresponding to $p_{z}$. Maybe
$n_{max}^{'}$ corresponding to $p_{z}~=~0$ is most meaningful to us, and so is
$p_{\rm F}(z)$ corresponding to $n~= 0$ (when $n = 0$, the spin is antiparallel
to $B$,and $\sigma~= -\frac{1}{2}$).

According to the definition of $E_{\rm F}(e)$ in Eq.(2), we obtain
$E_{\rm F}(e) \equiv p_{\rm F}(e)c$ if electrons are super-relativistic
($E_{\rm F}(e)\gg m_{e}c^{2}$). In the presence of a superhigh magnetic field
$B\gg B_{\rm cr}$, $E_{\rm F}(e) \gg m_{e}c^{2}$), hence we have the
following approximate relation
\begin{equation}
\nu_{max}(\sigma=-\frac{1}{2})^{'}=~\nu_{max}(\sigma=~\frac{1}{2})^{'}
\simeq Int[\frac{1}{2B^{*}}[(\frac{E_{\rm F}(e)}{m_{e}c^{2}})^{2}]~~.~~\label{14}
\end{equation}

From the analysis above, we can calculate the maximum of electron momentum
perpendicular to the magnetic field by
\begin{equation}
p_{\bot}^{2}(max)c^{2} =~2\nu_{max}^{'}2m_{e}c^{2}\mu_{e}^{'}B~~,~~
\label{15}
\end{equation}
where the relation of $2\mu_{e}^{'}B_{\rm cr}/m_{e}c^{2}= 1$ is used.
Inserting Eq.(14) into Eq.(15) gives
\begin{eqnarray}
&&p_{\bot}^{2}(max)c^{2} =~2\times \frac{1}{2B^{*}}(\frac{E_{\rm F}(e)}{m_{e}
c^{2}})^{2}\times 2m_{e}c^{2}\mu_{e}^{'}B \nonumber\\
&&\simeq B_{\rm cr}\times(\frac{E_{\rm F}(e)}{m_{e}c^{2}})^{2}\times
2m_{e}c^{2}\frac{e\hbar}{2m_{e}c} \nonumber\\
&&=\frac{m_{e}^{2}c^{3}}{e\hbar}\times(\frac{E_{\rm F}(e)}{m_{e}c^{2}})
^{2}\times 2m_{e}c^{2}\frac{e\hbar}{2m_{e}c}=~E_{\rm F}^{2}(e)~~.
~~\label{16}
\end{eqnarray}
Thus, we gain
\begin{eqnarray}
&&p_{\bot}(max)= p_{\rm F}(e)\simeq \frac{E_{\rm F}(e}{c}\nonumber\\
&&= 43.44\times(\frac{Y_{e}}{0.0535}\frac{\rho}{\rho_{0}}\frac{B}{B_{\rm cr}})
^{\frac{1}{4}}{\rm MeV/c} ~(B^{*} \geq 1~)~~.~~
\label{17}
\end{eqnarray}
As pointed out above, when $n= 0$, the electron Landau level is non-degenerate,
and $p_{z}$ has its maximum $p_{z}(max)$,
\begin{eqnarray}
&&p_{z}(max)= p_{\rm F}(e)\simeq \frac{E_{\rm F}(e)}{c}\nonumber\\
&&= 43.44\times(\frac{Y_{e}}{0.0535}\frac{\rho}{\rho_{0}}\frac{B}{B_{\rm cr}})
^{\frac{1}{4}}{\rm MeV/c}  ~(B^{*} \geq 1~)~~.~~
\label{18}
\end{eqnarray}
From Eq.(17) and Eq.(18), it's obvious that $p_{z}(max)=p_{\bot}(max)
= p_{\rm F}(e)$, which implies the electron pressure along the magnetic
field, $P_{z}(e)$, is equal to the electron pressure in the direction
perpendicular to the magnetic field, $P_{\bot}(e)$. The reason for this is
that in the interior of a magnetar, electrons are degenerate and super-relativistic,
and can be approximately treated as an ideal Fermi gas with equivalent pressures
in all directions, though the existence of Landau levels.  The equation of
$P_{e}$ in a superhigh magnetic field is consequently given by
\begin{eqnarray}
&&P_{e}= \frac{1}{3}\frac{2}{h^{3}}\int_{0}^{p_{\rm F}(e)}\frac{p_{e}^{2}c^{2}}
{(p_{e}^{2}c^{2}+ m_{e}^{2}c^{4})^{1/2}}4\pi p_{e}^{2}dp_{e}\nonumber\\
&&= \frac{m_{e}c^{2}}{\lambda_{e}^{3}}\phi(x_{e})= 1.412\times10^{25}\phi (x_{e})
 ~~{\rm dynes~cm^{-2}}~~,~~\label{19}
 \end{eqnarray}
where $\lambda_{e}=\frac{h}{m_{e}c}$ is the electron Compton wavelength,
and $\phi(x_{e})$ is the polynomial of a non-dimensional variable $x_{e}$
($x_{e}=\frac{p_{\rm F}(e)}{m_{e}c}\simeq \frac{E_{\rm F}(e)}
{m_{e}c^2}= 86.77 \times(\frac{\rho}{\rho_{0}}\frac{B}{B_{\rm cr}}\frac{Y_{e}}
{0.0535})^{\frac{1}{4}}$),
 \begin{equation}
 \phi(x_{e})=\frac{1}{8\pi^2}[x_{e}(1 + x_{e}^{2})^{\frac{1}{2}}
 (\frac{2x_{e}^{2}}{3}- 1)+ln[x_{e}+(1+x_{e}^{2})^{\frac{1}{2}}]]~~.~~\label{20}
\end{equation}
When $\rho\geq 10^{7}$~g~cm$^{-3}$, $x_{e}\gg 1$, and $\phi(x_{e})
\rightarrow \frac{x_{e}^{4}}{12\pi^2}$. Thus, Eq.(19) can be rewritten as
\begin{equation}
P_{e}\simeq 6.266\times10^{30}(\frac{\rho}{\rho_{0}}\frac{B}{B_{\rm cr}}
\frac{Y_{e}}{0.0535})~~{\rm dyne~cm^{-2}}~~.~~\label{21}.
\end{equation}

Comparing Eq.(21) with Eq.(7), it's easy to see  that $P_{e} \propto E_{\rm F}^{4}
(e)$. Employing Eq.(21), we plot the schematic diagrams of $P_{e}$
vs. $B$, as shown in Fig.1.

\begin{figure*}[htb]
\centerline{\psfig{file=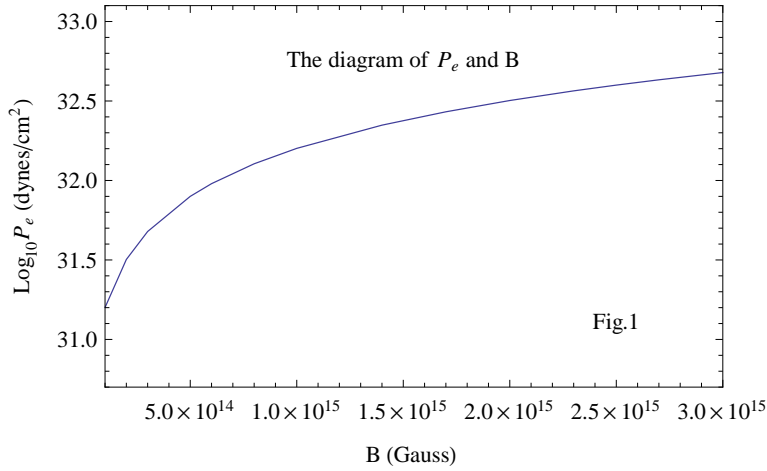,width=4.0in}}
\vspace*{8pt}
\caption{The electron pressure dependence of the magnetic field strength
when $\rho= \rho_{0}$ and $Y_{e}= 0.06$. When $B$ is assumed to be $(1.0
\times10^{14}\sim 3.0\times 10^{15}$)~G, $P_{e}$ is $(1.592\times
10^{31}\sim 4.776\times 10^{32}$~dynes~cm$^{-2}$), correspondingly.
\protect\label{fig1}}
\end{figure*}

From Fig.1, $P_{e}$ increases sharply with increasing $B$ when the
values of $Y_{e}$ and $\rho$ are given. We also present a schematic
illustration of $P_{e}$ as a function of $\rho$ in different
magnetic fields, as shown in Fig.2.
\begin{figure*}[htb]
\centerline{\psfig{file=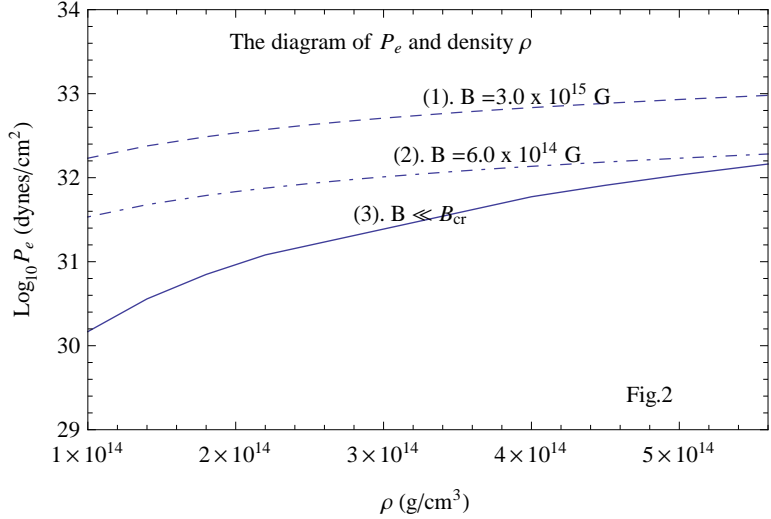,width=4.0in}}
\vspace*{8pt}
\caption{The electron pressure dependence of the matter density in different
magnetic fields when $Y_{e}= 0.06$. When $\rho$ is assumed to be $(1.0
\times 10^{14}\sim 5.6 \times10^{14}$)~g~cm$^{-3}$, the ranges of $P_{e}$
are $1.471\times 10^{30}\sim 1.454\times 10^{32}$~dynes~cm$^{-2}$,$3.411
\times10^{31}\sim 1.910\times 10^{32}$~dynes~cm$^{-2}$, and $1.521\times
10^{32}\sim 8.545\times 10^{32}$~dynes~cm$^{-2}$, respectively, corresponding
to $B^{*}\ll 1$, $B =~6.0 \times 10^{14}$~G and $B= 3.0\times 10^{15}$ G,
respectively. The solid line is obtained from Eq.(3) for the relativistic
electrons~(see Section 1).
\protect\label{fig2}}
\end{figure*}


As pointed in our previous study (see Ref.~\refcite{Gao11a}), the high Fermi
energy of electrons could be supplied by the release of the magnetic field
energy. Since $P_{e}$ is proportional to $E_{\rm F}^{4}(e)$, and
the later increases with increasing $B$, then the former also increases with
increasing $B$ naturally. Although we have presented a reasonable explanation
for high-value $P_{e}$ here, an important issue of electron distributions
among different Landau levels in a superhigh magnetic field remains unsolved.
According to a popular viewpoint that the stronger the magnetic field, the
smaller the maximum electron Landau level number $n_{max}$ becomes, the
number of electrons congregating in an exciting level ($n\geq 1$) decrease
with increasing $B$, if $B\gg B_{\rm cr}$, then $n_{max}=~1$ or 2, and the
overwhelming majority of electrons occupy in the ground level $n=~0$, which
causes a decrease in the momentum $p_{z}$, as well as a decrease in the
momentum $p_{\bot}$.  However, the Dirac Delta-function $\delta(\frac{p_
{\bot}}{m_{e}c}- [2(n +\sigma+ \frac{1}{2})B^{*}]^{\frac{1}{2}})$
obviously tells us that, given a Landau level number $n$, both the momentum $p_
{\bot}(n)\sim m_{e}c[2(n + \sigma + \frac{1}{2})B^{*}]^{\frac{1}{2}}$
and the magnetic energy $E_{B}(n)\sim p_{\bot}(n)c \sim m_{e}c^{2}[2(n
+ \sigma + \frac{1}{2})B^{*}]^{\frac{1}{2}}$ increase with increasing $B$,
which implies more electrons contribute to $p_{\bot}(n)$ and $E_{B}(n)$.
Thus, given a Landau level $n$, the number of electrons should increase with
increasing $B$, rather than decrease with $B$. With respect to $\Delta
p_{\bot}(n, n+1)$, the difference of $p_{\bot}$ between two adjacent landau
levels, when the Landau level number $n$ is given, $\Delta p_{\bot}(n, n+1)$
increases with increasing $B$; when $B$ is given, $\Delta p_{\bot}(n, n+1)$
decreases with increasing $n$, and if $n\rightarrow n_{max}$, then $\Delta
p_{\bot}(n, n+1)$ is too small to be taken into account. The increase of
$E_{\rm F}(e)$, together with the increase of number of electrons
populating in the vicinity of Fermi surface, can result in an increase in the
electron capture(EC) rate in the interior of a magnetar. The extreme activity
and instability of a magnetar may be explained by high-value $E_{\rm F}(e)$,
and the released thermal energy in the EC process could be the source of
magnetars' soft X/$\gamma$-ray radiations (see Ref.~\refcite{Gao12a},~\refcite{Gao11b},~\refcite{Gao12b}).

\section{The QED effects on equation of state}
A superhigh magnetic field is intriguing in the physical context because some
interesting phenomena would be expected from the quantum electrodynamic (QED)
effects. One of the QED phenomena is the vacuum polarization, which causes
large energy splitting of Landau levels, modifies the dielectric properties of
the medium, induces resonant conversion of photon modes (e.g., see Ref.~\refcite{Ventura79},
\refcite{Bulik97},~\refcite{Kohri02}),and decreases the equivalent width of the
ion cyclotron line making the line more difficult to observe (see Ref.~\refcite{Ho03}).

The observations display the presence of the hard X-ray spectra tails in magnetars,
recently detected by XMM, ASCA, RXTE and INTEGRAL (e.g.,~see Ref.~\refcite{Kuiper04},
~\refcite{Molkov05},~\refcite{Mereghetti05},~\refcite{Gotz06},~\refcite{Kuiper06},~\refcite
{den Hartog08}). Some authors have tried to interpret these observations by employing
Resonant Compton Scattering (RCS) by relativistic electrons in superhigh magnetic
fields (e.g., see Ref.~\refcite{Lyutikov06},~\refcite{Baring07},~\refcite{Rea08},~
\refcite{Nobili08},~\refcite{Baring11},~\refcite{Gonthier12}). RCS is also an
interesting phenomenon due to QED effects, and has been treated as a leading
emission mechanism for the hard X-ray radiation of magnetars. With respect to the
QED effects on the spin-down and heat evolution of magnetars, for details, see a
recent review of Ref.~\refcite{Battesti13}.

However, as an alternative, in this work, we shall concentrate on the QED effects on
different matter systems of a magnetar, and discuss the star's deformation due to strong
magnetic fields. This Section is composed of three sub-sections. For each subsection we
present different methods and considerations.
\subsection{The QED effects on the EOS of BPS model}
Considering shell effects on the binding energy of a given nucleus,
Salpeter (1961) first calculated the composition and EOS in the region of
$10^{7}-3.4\times10^{11}$~g~cm$^{-3}$~(see Ref.~\refcite{Salpeter61}). By
introducing the lattice energy, Baym, Pethick and Sutherland (hereafter
BPS model) improved on Salpeter's treatment, and described the nuclear
composition and EOS for catalyzed matter in complete thermodynamic
equilibrium below neutron drop $\rho_{d}\sim 4.4 \times 10^{11}$~g~cm$^{-3}$
energy (see~Ref.~\refcite{BPS71},~\refcite{Shapiro83}).

Here, we shall not discuss electrons in the low density regime $10^{4}\sim 10^{6}$~g
~cm$^{-3}$, because the electrons are nonrelativistic, and EOS may be treated via a
magnetic Thomas-Fermi (TF) type of model for a lattice of electrons and $^{56}_{26}Fe$
nuclei (see Ref.~\refcite{Fushiki89}). Since we are concerned about the role of magnetic
fields on EOS, a qualitative analysis of magnetic influence depending on the details of
the nuclear model will not be presented here. Also, we'll utilize the simple BPS model,
and no long consider other more complicated models (e.g., the model in Ref.~\refcite
{Haensel89}, in which the EOS is quite distinct from that of BPS model). We'll assume
an isotropic and homogenous matter pressure $P$ of this system. An anisotropy of the
total pressure $P_{tot}$ of the system due to a strong magnetic field will not be
discussed in this subsection.

According to BPS model, the matter energy density is given by
 \begin{equation}
\varepsilon = n_{N}(W_{N}(A, Z) + \varepsilon_{L}(Z, n_{e})+
\varepsilon_{e}(n_e)~~,~~\label{22}
\end{equation}
where $n_{N}$ is the number density of nuclei, $W_{N}(A, Z)$ is the mass-
energy per nucleus (including the rest mass of $Z$ electrons and $A$ nucleons);
$\varepsilon_{e}$ is the free electron energy including the rest mass of
electrons in a unit volume; $\varepsilon_{L}$ is the $bcc$ Coulomb lattice energy
per nucleus,
\begin{equation}
 \varepsilon_{L}= -1.444Z^{2/3}e^{2}e^{2}n_{e}^{4/3}~~ , ~~\label{23}
\end{equation}		
where the relations of $n_{N}=~n_{B}/A$ and $n_{e}=~Zn_{N}$ are used.
The matter pressure $p$ of the system is given by
\begin{equation}
 P= P_{e} + P_{L}= P_{e}+ \frac{1}{3}\varepsilon_{L}~~. ~~\label{24}
\end{equation}
In the case of the weak-field limit $B^{*}\ll 1$, $P_{e}$ in Eq.(24) is given by
Eq.(5). For a magnetic field $B^{*}\gg 1$, $P_{e}$ in Eq.(24) is given by
Eq.(21). For the specific equilibrium nuclei $(A, Z)$ in BPS model, we calculate the
values of quantities $\rho_{m}$, $n_{B}(m)$, $E_{\rm F}(e)$ and $P_{e}$,
as tabulated in Table 1.
\begin{table*}[htb]
\tbl{Values of $\rho_{m}$, $n_{B}(m)$, $E_{\rm F}(e)$ and $P_{e}$
in BPS model below neutron drip.}
{\begin{tabular}{@{}cccccccccc@{}} \toprule
Nuclei$^{\dag}$ & $Y_{e}$& $\rho_{m}^{0}$&$\rho_{m}^{1}$&$n_{B}^{0}(m)$&$n_{B}^{1}(m)$&$E_{\rm F}^{0}(e)$
&$E_{\rm F}^{1}(e)$&$P_{e}^{0}$&$P_{e}^{1}$\\
&         &g~cm$^{-3}$ & g~cm$^{-3}$ &cm$^{-3}$&cm$^{-3}$& MeV & MeV& dynes~cm$^{-2}$ &dynes~cm$^{-2}$ \\
\colrule
$^{56}_{26}Fe$ & 0.464 &7.99$\times 10^{6}$ &4.67$\times 10^{8}$ &4.812$\times 10^{30}$ &2.812$\times 10^{32}$ &0.95&8.77&7.141$\times 10^{23}$&9.069$\times 10^{27}$ \\
$^{62}_{28}Ni$ & 0.452 &2.71$\times 10^{8}$ &1.68$\times 10^{9}$ &1.632$\times 10^{32}$ &1.012$\times 10^{33}$ &2.61&12.01&7.556$\times 10^{25}$&3.174$\times 10^{28}$\\
$^{64}_{28}Ni$ & 0.437 &1.31$\times 10^{9}$ &2.78$\times 10^{9}$ &7.949$\times 10^{32}$ &1.674$\times 10^{33}$ &4.31&13.59&5.980$\times 10^{26}$&5.087$\times 10^{28}$\\
$^{66}_{28}Ni$ & 0.424 &1.54$\times 10^{9}$ &No$^{\ddag}$        &9.274$\times 10^{32}$ &No$^{\ddag}$          &4.45&No$^{\ddag}$&7.049$\times 10^{26}$&No$^{\ddag}$ \\
$^{86}_{36}Kr$ & 0.419 &3.11$\times 10^{9}$ &3.87$\times 10^{9}$ &1.873$\times 10^{33}$ &2.331$\times 10^{33}$ &5.66&14.49 &1.767$\times 10^{27}$&6.776$\times 10^{28}$\\
$^{84}_{34}Se$ & 0.405 &9.98$\times 10^{9}$ &1.14$\times 10^{10}$&6.009$\times 10^{33}$ &6.865$\times 10^{33}$ &8.49&18.84 &7.999$\times 10^{27}$&1.930$\times 10^{29}$\\
$^{82}_{32}Ge$ & 0.390 &2.08$\times 10^{10}$&2.10$\times 10^{10}$&1.252$\times 10^{34}$ &1.265$\times 10^{34}$ &11.44&21.75 &2.028$\times10^{28}$&3.428$\times 10^{29}$\\
$^{80}_{30}Zn$ & 0.375 &5.91$\times 10^{10}$&5.69$\times 10^{10}$&3.559$\times 10^{34}$ &3.426$\times 10^{34}$ &14.08&27.62 &7.741$\times10^{28}$&8.925$\times 10^{29}$\\
$^{78}_{28}Ni$ & 0.359 &8.21$\times 10^{10}$&8.13$\times 10^{10}$&4.944$\times 10^{34}$ &4.896$\times 10^{34}$ &20.01&28.86 &1.132$\times10^{29}$&1.221$\times 10^{30}$ \\
$^{126}_{44}Ru$& 0.349 &1.19$\times 10^{11}$&1.20$\times 10^{11}$&7.166$\times 10^{34}$ &7.226$\times 10^{34}$ &20.20&31.59 &1.789$\times10^{29}$&1.753$\times 10^{30}$\\
$^{124}_{42}Mo$& 0.339 &1.66$\times 10^{11}$&1.67$\times 10^{11}$&9.996$\times 10^{34}$ &1.006$\times 10^{35}$ &20.50&34.05 &2.678$\times10^{29}$&2.366$\times 10^{30}$\\
$^{122}_{40}Zr$& 0.328 &2.49$\times 10^{11}$&2.50$\times 10^{11}$&1.499$\times 10^{35}$ &1.506$\times 10^{35}$ &22.89&37.32 &4.404$\times10^{29}$&2.428$\times 10^{30}$\\
$^{120}_{38}Sr$& 0.317 &3.67$\times 10^{11}$&3.08$\times 10^{11}$&2.210$\times 10^{35}$ &2.216$\times 10^{35}$ &25.20&40.77 &7.052$\times10^{29}$&4.871$\times 10^{30}$\\
$^{122}_{38}Sr$& 0.311 &3.89$\times 10^{11}$&3.90$\times 10^{11}$&2.342$\times 10^{35}$ &2.348$\times 10^{35}$ &25.90&41.20 &7.455$\times10^{29}$&5.120$\times 10^{30}$\\
$^{118}_{36}Kr$& 0.305 &4.41$\times 10^{11}$&4.41$\times 10^{11}$&2.656$\times 10^{35}$ &2.656$\times 10^{35}$ &26.20&42.28 &7.805$\times10^{29}$&5.692$\times 10^{30}$\\
\botrule
\end{tabular}\label{ta1}}
\end{table*}

In this Table the signs of $`0'$ and $`1'$ denote $B^{*}\ll 1$ and $B^{*}= 100 $,
respectively, and $n_{B}(m)$ is the maximum equilibrium baryon number
density, corresponding to the maximum equilibrium density $\rho_{m}$ at
which the nuclide is present. The sign $`\dag'$ denotes that the first masses
are known experimentally (see Ref.~\refcite{Wapstra77}), and the remainder are
from the J\"{a}anecke-Gravey-Kelson mass formula (see Ref.~\refcite{Wapstra76}).
The sign $`\ddag'$ denotes that $^{62}_{28}Ni$ is found to be absent from the
equilibrium nucleus sequence, so is not presented. Furthermore, all digits in
Column 3 and Column 4 of this Table are cited from Lai and Shapiro (1991) (see
Ref.~\refcite{Lai91}). From Table 1, it's  obvious that the electron pressure
$P_{e}$, as well as $E_{\rm F}(e)$, increases with matter density and
magnetic field, and a strong magenetic field alters the nucleus transition
densities for the low-$A$ nuclei. In a strong magnetic field, as the density
increases, the nuclei become increasingly saturated with neutrons, however,
neutron drip occurs at the same density because we adopts a nearly constant
minimum Gibbs free energy per nucleon, $g_{min}=m_{n}c^{2}$.  From
Table 1, the calculated values of $P_{e}$ are in the range of $9.069
\times 10^{27}\sim 5.692\times 10^{30}$~dynes~cm$^{-2}$ corresponding to a
density range of $4.67\times 10^{8}\sim 4.41 \times 10^{11}$ when $B^{*}=~100$.
Based on this Table we plot four schematic sub-diagrams of QED effects on the
EOS, as shown in Fig.3.

\begin{figure*}[htb]
\begin{center}
\begin{tabular}{cc}
\scalebox{0.78}{\includegraphics{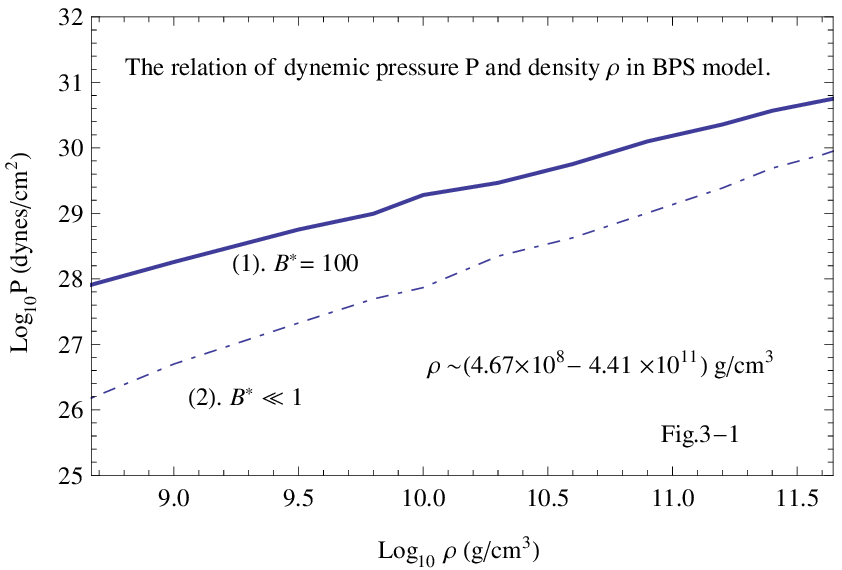}}&\scalebox{0.78}{\includegraphics{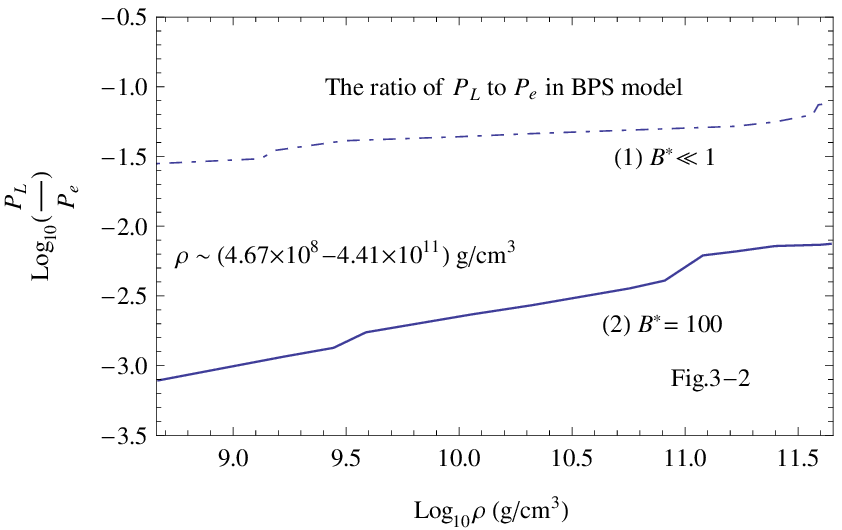}}\\
(a)&(b)\\
\scalebox{0.78}{\includegraphics{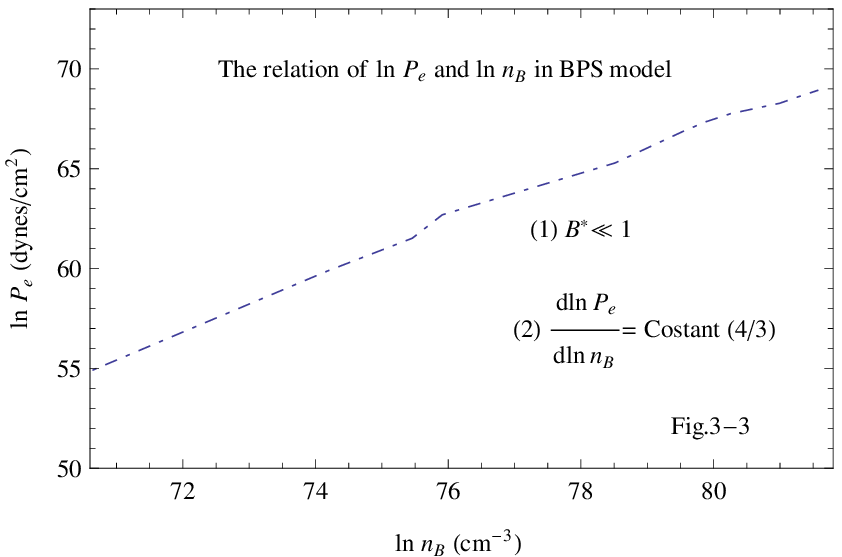}}&\scalebox{0.78}{\includegraphics{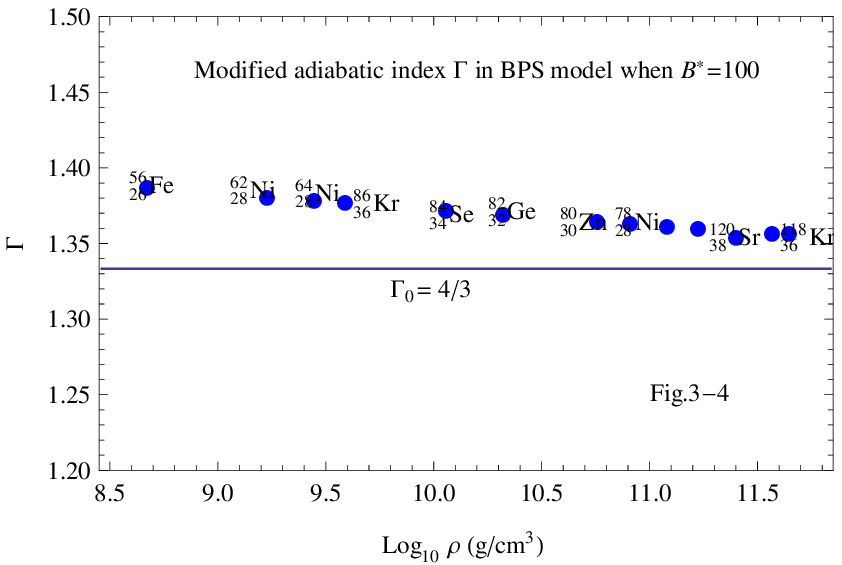}}\\
(c)&(d)\\
\end{tabular}
\end{center}
\caption{The QED effects on the EOS of BPS model below neutron drip.
Sub-diagrams Fig.3-1, Fig.3-2, Fig.3-3 and Fig.3-4 give $P$ versus $\rho$, $P_{L}/P_{e}$
versus $\rho$, $\ln P_{e}$ versus $\ln n_{B}$ and $\Gamma$ versus $\rho$ for
specific nucleus, respectively. Here, $\rho$ denotes the maximum equilibrium
density, $\Gamma$ is the modified adiabatic index (e.g., unlike the adiabatic
index $\Gamma_{0}$ in the weak-field limit, $\Gamma$ is an adiabatic index after
taking into account magnetic effects.
\protect\label{fig3}}
\end{figure*}

From the sub-diagram Fig.3-1, the matter pressure $P$ described by Eq.(24)
increases obviously with density $\rho$ for both in a superhigh magnetic field
and in the weak-field limit. Furthermore, given a density $\rho$, the value of
$P$ in the case of the former ($B^{*}= 100$)is bigger than that of $P$ in
the case of the latter ($B^{*}\ll 1$), which can be easily seen from the
curves of Fig.3-1. From the sub-diagram Fig.3-2, the ratios of $P_{L}$ to
$P_{e}$ are about $10^{-2}$ and $10^{-3}$, respectively, corresponding to $B^{*}
= 100$ and $B^{*}\ll 1$, respectively. Therefore, one can infer that the contribution
of $P_{L}$ to the matter pressure $P$ can be ignored if $B^{*}\geq 100$. In the
sub-diagram Fig.3-3, we fit a curve of $\ln P_{e}$ versus $\ln n_{B}$ in the
weak-field limit. Since the electrons are relativistic, the adiabatic index
$\Gamma$ (i.e., the slope of the curve) is a constant of $4/3$, which is the
standard relativistic value. In a strong magnetic field $B^{*}= 100$, due to the
changes of $P_{\rm e}$ and $\rho_{m}$, the values of modified $\Gamma$ are
slightly higher than the constant of $4/3$, i.e, $\Gamma\equiv \ln P_{e}/\ln n_{B}
\simeq 1.36\sim 1.39$, as illustrated in the sub-diagram Fig.3-4.

\subsection{The QED effects on the EOS of BBP model}
In the domain above neutron drip, as the matter density increases and the
nuclei become more neutron rich, and the system becomes a mixture of nuclei,
free neutrons, and electrons. When a critical value of baryon number density
is reached, the nuclei disappear, essentially by merging together. Here we
shall focus on the EOS of Baym, Bethe, and Pethick (1971) (see Ref.~
\refcite{BBP71})(hereinafter BBP), which describes such a system more
successfully, compared with other models.

In this subsection, we'll also assume an isotropic and homogenous matter
pressure $P$ of this system, and will not discuss an anisotropy of the total
pressure $P_{tot}$ of the system due to a strong magnetic field.

In the work by BBP, the major improvement was the introduction of a
compressible liquid drop model of the nuclei (see Ref.~\refcite{BBP71}).
The energy density of BBP model is written as
\begin{equation}
\varepsilon= \varepsilon_{e}(n_{e})+ n_{N}(W_{N} + W_{L})
+ n_{n}(1-~V_{N}n_{N})W_{n}~~,  ~~ \label{25}
\end{equation}
where $n_{n}$ is the number density of neutrons outside of nuclei
(hereinafter neutron gas), and the new feature is the dependence on the
volume of a nucleus $V_{N}$, which decreases with the outside pressure
of the neutron gas. The baryon number density in this model is
\begin{equation}
n_{B}= An_{N} + (1- V_{N}n_{N})n_{n}~~~,~~\label{26}
 \end{equation}
where $V_{N}n_{N}$ and $1- V_{N}n_{N}$ are the fraction of
volume occupied by nuclei, and the fraction occupied by the neutron gas, respectively.
The matter pressure $p$ in BBP model is given by
\begin{equation}
P= P_{n} + P_{e} + P_{L}~~.~~\label{27}
 \end{equation}
In the original work of BBP (1971), the three terms of $P_{n}$, $P_{e}$,
$P_{L}$ in Eq.(27) are calculated by $P_{n}=n_{n}^{2}\frac{\partial
E_{n}}{\partial n_{n}}$, $P_{e}=n_{e}^{2}\frac{\partial E_{e}}
{\partial n_{e}}$, and~$P_{L}=n_{N}^{2}\frac{\partial E_{L}}
{\partial n_{N}}$, respectively. Ignoring the details of calculations, we
list the values of chemical potentials $\mu_{e}$ and $\mu_{n}$ in Table 2,
and cite the results of $P_{n}$, $P_{e}$ and $P$ in plotting Fig.4 from
BBP (1971) and the review of Canuto (1974)(see Ref.~\refcite{Canuto74}).

In a superhigh magnetic field, according to our model, the three pressure
terms in Eq.(27) are approximately treated as follows: Sice the ratio of
$P_{L}/P$ is $\sim 10^{-3}-10^{-4}$, the contribution of negative lattice
pressure $P_{L}$ to the total dynamic pressure $P$ can be neglected; As in
the magnetic BPS model, the electron pressure $P_{e}$ is also given by
Eq.(21); In order to calculate the neutron pressure $P_{n}$, at first,
ignoring the neutrino chemical potential $\mu_{\nu}$, we determine $\mu_{n}$
via $\beta$-equilibrium under a uniform superhigh magnetic field,
\begin{equation}
\mu_{e} + \mu_{p} + m_{p}c^{2} = \mu_{n} + m_{n}c^{2}~~~, ~~\label{28}
 \end{equation}
where $\mu_{e}$ is the electron chemical potential, i.e., the electron Fermi
energy $E_{\rm F}(e)$, including the rest-mass energy $m_{e}c^{2}$,
$\mu_{p}$ is the proton chemical potential, i.e., the proton Fermi energy,
not including the rest-mass energy $m_{p}c^{2}$ (notice that in BBP model the protons do
not contribute the matter pressure because $\mu_{p}$ is negative),
and $\mu_{n}$ is also called the neutron Fermi energy $E_{\rm F}^{'}(n)$,
not including its rest-mass energy $m_{n}c^{2}$, $\mu_{n}=E_{\rm F}^{'}
(n)=~\frac{p_{\rm F}^{2}(n)}{2m_{n}}$, secondly, by defining a
non-dimensional variable $x_{n}=\frac{p_{\rm F}(n)}{m_{n}c}= \frac
{\sqrt{2m_{n}\mu_{n}}}{m_{n}c}=\sqrt{\frac{2\mu_{n}}{m_{n}c^2}} $, we
gain the polynomial $\phi(x_{n})$ of the variable $x_{n}$,
 \begin{equation}
 \phi(x_{n})= \frac{1}{8\pi^2}[x_{n}(1 + x_{n}^{2})^{\frac{1}{2}}(\frac{2x_{n}^{2}}{3}- 1)+
 ln[x_{n}+(1+x_{n}^{2})^{\frac{1}{2}}]]~~, ~~\label{29}
\end{equation}
where the neutrons are nonrelativistic (the density $\rho\ll 6\times 10^{15}$~g~cm$^{-3}$
in BBP model), then $x_{e}\ll 1$, and $\phi(x_{n})\rightarrow \frac{1}
{15\pi^2}(x_{n}^{5}- \frac{5}{14}x_{n}^{7}+\frac{5}{24}x_{n}^{9})$;
finally, following Shapiro \& Teukolsky (1983)(hereafter ST, see Ref.~\refcite
{Shapiro83}), we can calculate the value of $P_{n}$ by
\begin{equation}
 P_{n}=\frac{m_{n}c^{2}}{\lambda_{n}^{3}}\phi (x_{n})~=
 1.624 \times 10^{38}\frac{1}{15\pi^2}(x_{n}^{5}-\frac{5}{14}x_{n}
 ^{7}+~\frac{5}{24}x_{n}^{9}) ~{\rm dynes~cm^{-2}}~~, ~~\label{30}
 \end{equation}
where $\lambda_{n}~=~\frac{h}{m_{n}c}$ is the Compton wavelength of a neutron. In
Table 2 we present the values of $\mu_{e}$ and $\mu_{n}$ corresponding to
different magnetic fields in BBP model.

\begin{table*}[htb]
\tbl{Values of $\mu_{e}$ and $\mu_{n}$ in magnetic BBP model.}
{\begin{tabular}{@{}cccccccccccc@{}} \toprule
$\rho$ & $A$ & $Z$ & $n_{N}\times 10^{-6}$& $n_{B}$ &$Y_{e}^{\dag}$&$\mu_{e}^{0}$& $\mu_{e}^{1}$& $\mu_{e}^{2}$
&$\mu_{n}^{0}$& $\mu_{n}^{1}$& $\mu_{n}^{2}$\\
(g~cm$^{-3}$) & & &  (fm$^{-3}$)& (cm$^{-3}$)& &(MeV) &(MeV)&(MeV) &(MeV)& (MeV) &(MeV) \\
\colrule
4.66$\times 10^{11}$&127 &40 &2.02 &2.806$\times 10^{35}$  &0.2879 &26.31 &35.54 &42.26 &0.14 &9.37 &16.09  \\
6.61$\times 10^{11}$&130 &40 &2.13 &3.981$\times 10^{35}$  &0.2315 &26.98 &36.72 &43.67 &0.37 &10.11 &17.06 \\
8.79$\times 10^{11}$&134 &41 &2.23 &5.293$\times 10^{35}$  &0.1727 &27.51 &36.86 &43.83 &0.55 &9.90 &16.87 \\
1.20$\times 10^{12}$&137 &42 &2.34 &7.226$\times 10^{35}$  &0.1360 &28.13 &37.32 &44.38 &0.75 &9.94 &17.10 \\
1.47$\times 10^{12}$&140 &42 &2.43 &8.852$\times 10^{35}$  &0.1153 &28.58 &37.67 &44.80 &0.91 &10.10 &17.13 \\
2.00$\times 10^{12}$&144 &43 &2.58 &1.204$\times 10^{36}$  &0.0921 &29.33 &38.47 &45.74 &1.15 &10.29 &17.56 \\
2.67$\times 10^{12}$&149 &44 &2.74 &1.608$\times 10^{36}$  &0.0749 &30.15 &39.26 &46.69 &1.42 &10.53 &17.96  \\
3.51$\times 10^{12}$&154 &45 &2.93 &2.114$\times 10^{36}$  &0.0624 &31.05 &40.17 &47.76 &1.71 &10.83 &18.42\\
4.54$\times 10^{12}$&161 &46 &3.14 &2.734$\times 10^{36}$  &0.0528 &32.02 &41.08 &48.86 &2.01 &11.07 &18.85 \\
6.25$\times 10^{12}$&170 &48 &3.45 &3.764$\times 10^{36}$  &0.0439 &33.43 &42.52 &50.56 &2.45 &11.54 &19.58 \\
8.38$\times 10^{12}$&181 &49 &3.82 &5.046$\times 10^{36}$  &0.0371 &34.98 &43.84 &52.14 &2.91 &11.71 &20.07  \\
1.10$\times 10^{13}$&193 &51 &4.23 &6.624$\times 10^{36}$  &0.0326 &36.68 &45.44 &54.03 &3.41 &12.17 &20.76  \\
1.50$\times 10^{13}$&211 &54 &4.84 &9.033$\times 10^{36}$  &0.0289 &39.00 &47.64 &56.65 &4.07 &12.71 &21.72 \\
1.99$\times 10^{13}$&232 &57 &5.54 &1.198$\times 10^{37}$  &0.0264 &41.56 &49.99 &59.45 &4.77 &13.20 &22.66\\
2.58$\times 10^{13}$&257 &60 &6.36 &1.554$\times 10^{37}$  &0.0246 &44.37 &52.41 &62.32 &5.51 &13.55 &23.46  \\
3.44$\times 10^{13}$&296 &65 &7.52 &2.071$\times 10^{37}$  &0.0236 &48.10 &55.73 &66.28 &6.47 &14.10 &24.65  \\
4.68$\times 10^{13}$&354 &72 &9.12 &2.818$\times 10^{37}$  &0.0233 &52.95 &59.99 &71.38 &7.67 &14.71 &26.10 \\
5.96$\times 10^{13}$&421 &78 &10.7 &3.589$\times 10^{37}$  &0.0235 &57.56 &64.38 &76.56 &8.77 &15.90 &27.77 \\
8.01$\times 10^{13}$&548 &89 &13.1 &4.824$\times 10^{37}$  &0.0242 &64.32 &69.25 &82.35 &10.36 &15.29 &28.39  \\
9.83$\times 10^{13}$&683 &100 &15.0 &5.920$\times 10^{37}$ &0.0253 &69.81 &73.76 &87.72 &11.66 &15.61 &29.57  \\
1.30$\times 10^{14}$&990 &120 &17.8 &7.828$\times 10^{37}$ &0.0233 &78.58 &80.57 &95.82 &13.77 &15.76 &30.01  \\
1.72$^{\ddag}\times 10^{14}$&1640 &157 &19.6 &1.035$\times10^{38}$&0.0297 &88.84 &88.28 &104.98&16.39 &15.83 &32.53  \\
2.00$\times 10^{14}$&2500 &210 &18.8 &1.205$\times 10^{38}$&       &      &      &      &      &      &      \\
2.26$\times 10^{14}$&4330 &290 &15.4 &1.361$\times 10^{38}$&       &      &      &      &      &      &      \\
2.39$\times 10^{14}$&7840 &445 &11.0 &1.439$\times 10^{38}$&       &      &      &      &      &      &     \\
\botrule
\end{tabular}\label{ta2}}
\end{table*}

In this Table, the signs of $`0'$, $`1'$ and $`2'$ denote $B^{*}\ll 1$,
$B^{*}= 50 $ and $B^{*}= 100 $, respectively, and the data of columns 1, 2, 3,
4, 7 and 10 are from Table 2 of Canuto (1974)(see Ref.~\refcite{Canuto74}). The
sign `$\dag$' denotes the electron fraction $Y_{e}$ is given by $Y_{e}
=Y_{p}= \frac{Zn_{N}}{n_{B}}\simeq \frac{Zn_{N}}{\rho/m_{u}}$, where
$m_{u}= 1.66057\times 10^{-24}$~g~is the atom-mass unit because $n_{B}$
always can be treated as an invariable quantity in spit of a strong magnetic
field. The sign `$\ddag$'denotes that the EOS of BBP model has been criticized
by some authors (e.g., see Ref.~\refcite{Canuto74},~\refcite{Shapiro83}),
due to a monotonic and arbitrary increase of $Z$ with $A$, which causes the
relation of $P$ and $\rho$ to be changed not much, so we stop listing the
related calculations in the higher density region, $\rho\geq 1.72 \times
10^{14}$~g~cm$^{-3}$.

The values of $\mu_{n}$ are obtained from Eq.(28) ignoring the changes
of kinetic energies and potential energies of protons in nuclei and the
transformations between protons and neutrons caused by strong magnetic
fields (i.e., keeping $\mu_{p}$ and $Y_{p}$ or $Y_{e}$ constant).
In a superhigh magnetic field, the system under consideration is
still a mixture of free neutrons, electrons, and nuclei. Since there is a
lack of experimental data on the changes and transformations above, we still
take the quantities $\mu_{p}$ and $Y_{p}$ as constants approximately
when calculating $\mu_{e}$ and $\mu_{n}$. If $B\gg 10^{20}$ G, these
ignorers are not permitted due to larger calculating errors. Note that such
strong magnetic fields will no longer be taken into account according
to our magnetar model (for details, see the following subsection). However,
these ignorers can not affect our main result here that the values of $\mu_{e}$
and $\mu_{n}$ increase with increasing magnetic fields and densities.  Based
on Table 2 we plot four schematic sub-diagrams of QED effects on the EOS of BBP model,
as shown in Fig.4.

\begin{figure*}[htb]
\begin{center}
\begin{tabular}{cc}
\scalebox{0.78}{\includegraphics{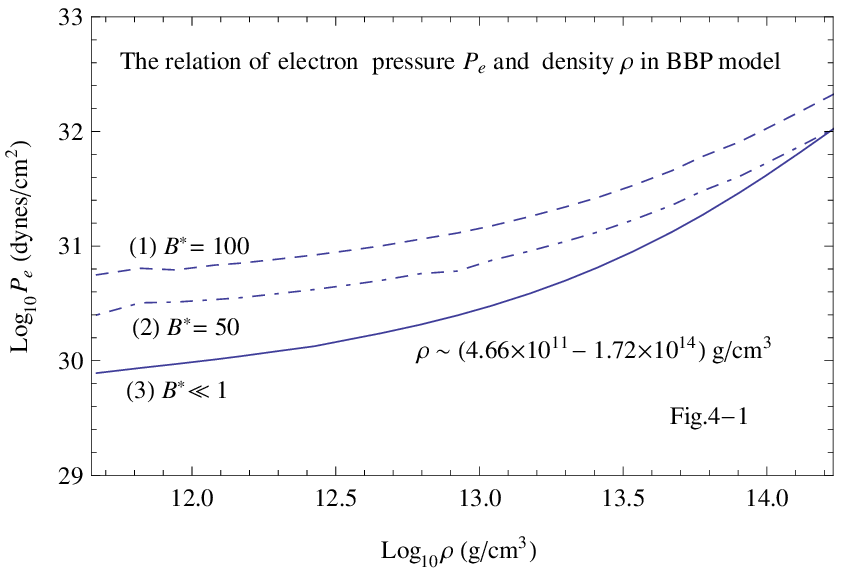}}&\scalebox{0.78}{\includegraphics{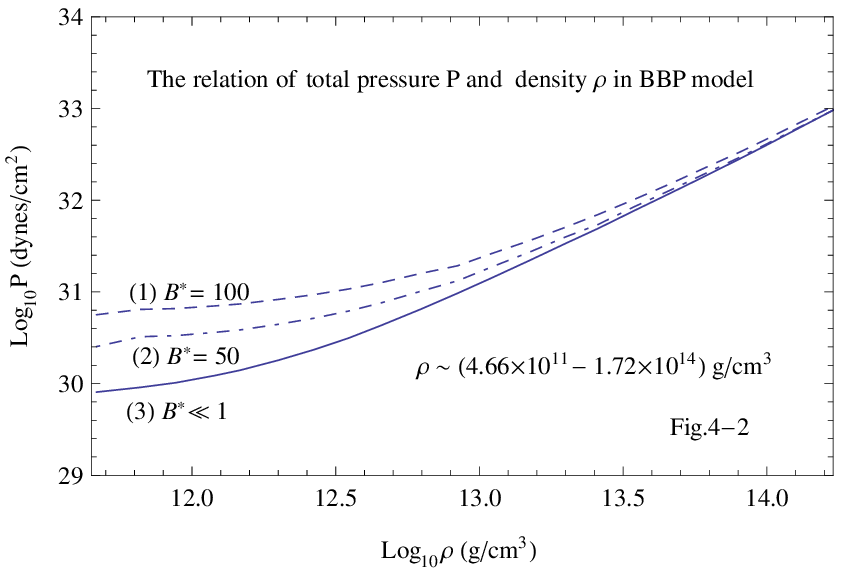}}\\
(a)&(b)\\
\scalebox{0.78}{\includegraphics{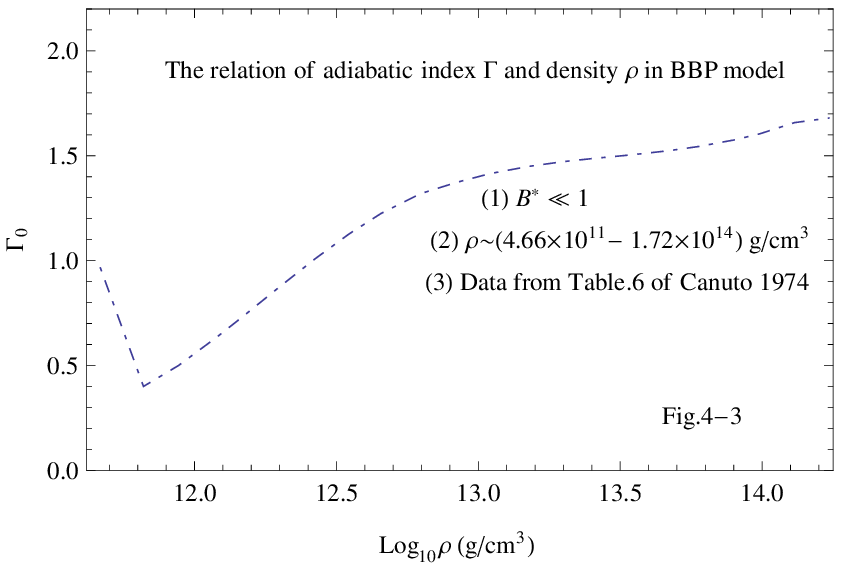}}&\scalebox{0.78}{\includegraphics{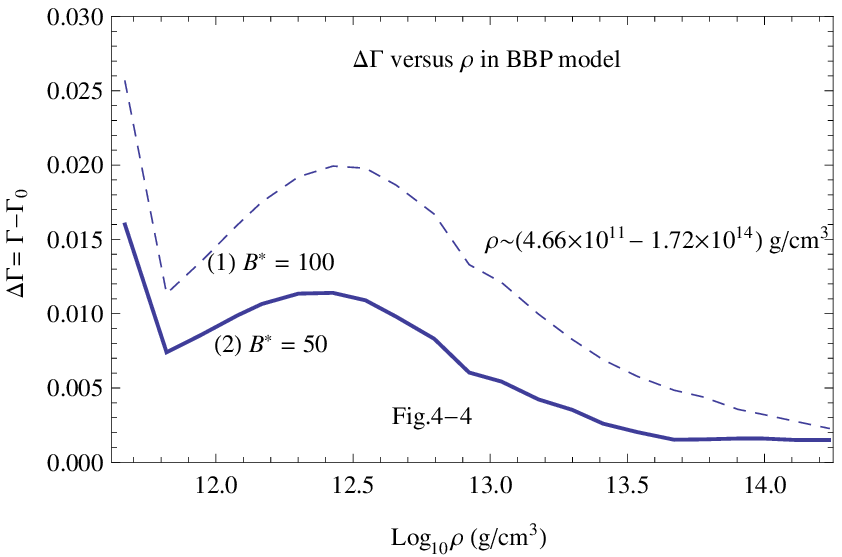}}\\
(c)&(d)\\
\end{tabular}
\end{center}
\caption{The QED effects on the EOS of BBP model. Sub-diagrams Fig.4-1, Fig.4-2,
Fig.4-3 and Fig.4-4 give $P_{e}$ versus $\rho$, $P$ versus $\rho$,
$\Gamma$ versus $\rho$, and $\Delta \Gamma$ versus $\rho$, respectively,
for this EOS in different magnetic fields. The matter pressure $P$ of this system is
described by Eq.(27), assuming that $\mu_{p}$ and $Y_{e}$ are
invariant approximately.
 \protect\label{fig4}}
\end{figure*}

From sub-diagrams Fig.4-1 and Fig.4-2, both $P_{e}$ and $P$ obviously
increase with density $\rho$ for the three cases. As in BPS model above,
given a same density $\rho$, the stronger the magnetic field is, the higher
the values of $P_{e}$ and $P$  become. However, the increments of
$P_{e}$ and $P$ in the lower density region are larger than those in the
higher density region. The main reason for this is that the increments of
quantities $\mu_{e}$ and $\mu_{n}$ in the lower density region are
larger than those in the higher density region, which can be easily seen in
Table 2. In Fig.4-3 the curve of adiabatic index $\Gamma$ and $\rho$ in the
weak-field limit is obtained by fitting with the data of Table 6 in Canuto
(1974)~(see Ref.~\refcite{Canuto74}). The key feature of the curve in this
sub-diagram is that $\Gamma$ decreases with increasing $\rho$ firstly, and
then increases with increasing $\rho$. This phenomenon can be explained as
follows: Firstly, as the neutron drip density $\rho_{d}$ is approached, the
electrons are extremely relativistic, and the matter  pressure is almost
entirely due to electrons, therefore $\Gamma_{0}= 4/3$; Secondly, slightly
above $\rho_{d}$, the low-density neutron gas contributes appreciably to
the density but not much to the matter pressure, and thus $\Gamma$ falls
sharply. This drop in BBP model is described as $\Gamma=\frac{4}{3}[1-a(\rho-
\rho_{d})^{1/2}]$, where $a$ is a positive constant. Thirdly, as the density
increases (about $\rho\geq 10^{12}$~g~cm$^{-3}$), the free neutrons
nevertheless contribute to an increasingly larger fraction of the matter
pressure. According to our calculations, in a strong magnetic field, e.g.,
$B^{*}\geq 50$, when $\rho \sim 10^{12}$~g~cm$^{-3}$, $P_{\rm n}/P
\sim 0.1-0.2$, while $\rho \sim 10^{13}-10^{14}$~g~cm$^{-3}$, then $P_{n}/P
\sim 0.8 -0.99$, assuming that $\rho_{d}$ is unchanged. In addition, $\Gamma$
grows with increasing $B$ and $\rho$ slightly ($\Delta \Gamma \sim
10^{-2}-10^{-3}$) due to an increase in $P$, as shown in Fig.4-4.
\subsection{The QED effects on the EOS of ideal $npe$ gas}
In this subsection, we consider a homogenous ideal $npe$ gas under $\beta$-
equilibrium, and adopt ST approximation corresponding to the weak-field limit
(see Ref.~\refcite{Shapiro83}) as the main method to treat EOS of this system
in the density range of $0.5\rho_{0}\sim 2\rho_{0}$ where electrons are relativistic,
neutrons and protons are non-relativistic. According to ST, when $\rho\gg
10^{13}$~g~cm$^{-3}$, neutrons dominate in the interior of a NS, $\rho
\approx m_{n}n_{n}$, then $n_{n}=1.7\times 10^{38}(\frac{\rho}
{\rho_{0}})$~cm$^{-3}$ (see Ref.~\refcite{Shapiro83}); employing $\beta$-
equilibrium and charge neutrality gives $n_{p}= n_{e}= 9.6\times
10^{35}(\frac{\rho}{\rho_{0}})^{2}$~cm$^{-3}$; $\beta$-equilibrium implies
energy conservation and momentum conservation ($p_{\rm F}(p)=p_{\rm F}
(e)$), we get $E_{\rm F}(e)= \mu_{n}= E_{\rm F}^{'}(n)
=p_{\rm F}^{2}(n)/2m_{n}=60(\frac{\rho}{\rho_{0}})^{2/3}$~MeV, and
$\mu_{p}=E_{\rm F}^{'}(p)=p_{\rm F}^{2}(p)/2m_{p}= 1.9
(\frac{\rho}{\rho_{0}})^{4/3}$~MeV; the isotropic matter pressure $P$ is given by
\begin{eqnarray}
&&P= P_{e}+ P_{p}+ P_{n}\nonumber\\
&&= \frac{m_{e}c^{2}}{\lambda_{e}^{3}}\phi(x_{e})+ \frac{m_{p}c^{2}}{\lambda_{p}^{3}}\phi(x_{p})+
\frac{m_{n}c^{2}}{\lambda_{n}^{3}}\phi(x_{n})~~, ~~\label{31}
 \end{eqnarray}
where $x_{p}=~\frac{p_{\rm F}(p)}{m_{p}c}=
\frac{\sqrt{2m_{p}\mu_{p}}}{m_{p}c}~=~\sqrt{\frac{2\mu_{p}}{m_{p}c^2}}$, the
expression of $\phi(x_{p})$ is completely similar to that of $\phi(x_{\rm n})$ (see Eq.(29)).

Based on the above results, we gain the following useful formulae:
\begin{eqnarray}
&&P_{p}=1.169\times 10^{30}(\frac{\rho}{\rho_{0}})^{\frac{10}{3}}~~{\rm dynes~cm^{-2}}~~,\nonumber\\
&&P_{e}=1.825\times 10^{31}(\frac{\rho}{\rho_{0}})^{\frac{8}{3}}~~{\rm dynes~cm^{-2}}~~,\nonumber\\
&&P_{n}=6.807\times 10^{33}(\frac{\rho}{\rho_{0}})^{\frac{5}{3}}~~{\rm dynes~cm^{-2}}~~,\nonumber\\
&& Y_{e}=\frac{n_{e}}{n_{p}+n_{n}}\simeq \frac{n_{e}}{n_{n}}= 0.005647(\frac{\rho}{\rho_{0}})~~.~~\label{32}
 \end{eqnarray}
Be note that these formulae in Eq.(32) always hold approximately in an ideal $npe$
gas when $B^{*}\ll 1$ and $\rho\sim 0.5\rho_{0}-2\rho_{0}$.

Our methods to treat EOS of an ideal $npe$ gas (system) under $\beta$-equilibrium
in superhigh magnetic fields are introduced as follows: Combining Eq.(7) with momentum
conservation $p_{\rm F}(p)= p_{\rm F}(e)$ gives the chemical potential $\mu_{p}= E_{\rm F}
^{'}(p)= 1.005(\frac{B}{B_{\rm cr}}\frac{\rho}{\rho_{0}}\frac{Y_{e}}{0.0535})^{\frac{1}{2}}$~
MeV, and the non-dimensional variable $x_{p}= \sqrt{\frac{2\mu_{p}}{m_{p}c^2}}\simeq 4.626
\times 10^{-2}(\frac{B}{B_{\rm cr}}\frac{\rho}{\rho_{0}}\frac{Y_{e}}{0.0535})^{\frac{1}{4}}$;
From Eq.(28), we get the non-dimensional variable,
\begin{eqnarray}
 &&x_{n}= \sqrt{\frac{1}{m_{n}c^2}}(2\times(43.44(\frac{B}{B_{\rm cr}}\frac{\rho}{\rho_{0}}\frac{Y_{e}}{0.0535})^{1/4}~\nonumber\\
 &&-1.29 + 1.005(\frac{B}{B_{\rm cr}}\frac{\rho}{\rho_{0}}\frac{Y_{e}}{0.0535})^{1/2}))^{1/2}. ~~~\label{33}
 \end{eqnarray}
 Thus, re-solving Eq.(31) gives the expression for
the isotropic matter pressure $P$,
\begin{eqnarray}
&&P= \frac{m_{e}c^{2}}{\lambda_{e}^{3}}\phi(x_{e})+ \frac{m_{p}c^{2}}{\lambda_{p}^{3}}\phi(x_{p})+
\frac{m_{n}c^{2}}{\lambda_{n}^{3}}\phi(x_{n})\nonumber\\
&&= 6.266\times 10^{30}(\frac{\rho}{\rho_{0}}\frac{B}{B_{\rm cr}}\frac{Y_{e}}{0.0535})+2.324\times 10^{26}
(\frac{\rho}{\rho_{0}}\frac{B}{B_{\rm cr}}\frac{Y_{e}}{0.0535})^{\frac{5}{4}}\nonumber\\
&&+ 1.624 \times 10^{38}\frac{1}{15\pi^2}(x_{n}^{5}-\frac{5}{14}x_{n}^{7}+\frac{5}{24}x_{n}^{9})~{\rm dyne~cm^{-2}}~~,~~\label{34}
 \end{eqnarray}
where $x_{n}$ is determined by Eq.(33). The above equation always approximately hold in an ideal $npe$
gas when $B^{*}\gg 1$ and $\rho~\sim 0.5\rho_{0}-~2\rho_{0}$.

In order to calculate the matter pressure $P$ of an ideal $npe$ gas in
a superhigh magnetic field, it's better to an analytical expression for
$Y_{p}$ (or $Y_{e}$) and $B$. Unfortunately, so far such an expression
for $Y_{e}$ and $B$ has not been obtained yet. Furthermore, some related
researches are unauthentic, due to the lack of observational supports.
For example, Chakrabarty et~al.(1997) studied the gross properties of
cold symmetric matter, and matter in $\beta-$equilibrium under the
influences of superhigh magnetic fields using a relativistic Hartree
theory (see Ref.~\refcite{Chakrabarty97}). Their main conclusions are as
follows: Superhigh magnetic fields $\sim 10^{20}$ G could exist in the
interior of a NS; $Y_{e}$ (or $Y_{p}$) is a strong function of $B$ and
$\rho$; when an intense magnetic field is approached to the proton
critical magnetic field, $B_{\rm cr}^{p}\sim 1.48\times 10^{20}$~G, the
value of $Y_{p}$ (or $Y_{e}$) is expected to be enhanced considerably;
by strongly modifying the phase spaces of protons (electrons), the field
of $B\sim10^{20}$ G can bring on a substantial $n\rightarrow p$ conversion,
and the system is hence transformed into highly proton-rich matter with
distinctively softer EOS (see Ref.~\refcite{Chakrabarty97}). However,
up to date, there have been no observations indicating the existence of
fields $B\geq 10^{17}$ G inside a NS as mentioned in Section 1 of this
paper.  To sum up, magnetic fields of such magnitude ($\sim 10^{20}$ G)
inside NSs are unauthentic, and are not consistent with our magnetar
model (see Ref.~\refcite{Peng07}, Ref.~\refcite{Gao11a}, Ref.~
\refcite{Gao11b}). In this letter, we assume the maximum magnetic field
of a magnetar, $B\sim(3-4)\times 10^{15}$ G, in which $n_{B}$ always
keeps invariable, and the changes of $n_{p}$ (or $n_{e}$) and $n_{n}$
caused by the equilibrium process of $e^{-}+p \leftrightarrow n+ \nu$
are too small to be considered. Thus, the magnetic effects on the
proton (or electron) fraction also can be ignored. In Table 3 we
present the values of $E_{\rm F}(e)$, $\mu_{p}$, $\mu_{n}$ and $P$ in
ideal $npe$ gas corresponding to two cases: $B^{*}\ll 1$ and $B^{*}= 100 $.
\begin{table*}[htb]
\tbl{Values of $E_{\rm F}(e)$, $\mu_{p}$, $\mu_{n}$ and $P$ in ideal
$n-p-e$ gas corresponding to $B^{*}\ll 1$ and $B^{*}= 100 $.}
{\begin{tabular}{@{}ccccccccc@{}} \toprule
$\rho$ & $E_{\rm F}^{0}(e)$& $\mu_{p}^{0}$ &$\mu_{n}^{0}$&$E_{\rm F}^{1}(e)$
&$\mu_{\rm p}^{1}$ &$\mu_{\rm n}^{1}$&$P^{0}$&$P^{1}$\\
(g~cm$^{-3}$) &(MeV) &(~MeV)&(~MeV~) &(~MeV~)& (~MeV~) &(~MeV~)&(~dynes~cm$^{-2}$)&(~dynes~cm$^{-2}$) \\
\colrule
1.4$\times 10^{14}$&37.8 &0.8 &37.8 &55.4 &1.6 &55.7 &2.147$\times 10^{33}$ &5.247$\times 10^{33}$ \\
1.8$\times 10^{14}$&44.7 &1.1 &44.7 &62.8 &2.1 &63.6 &3.265$\times 10^{33}$ &7.189$\times 10^{33}$   \\
2.0$\times 10^{14}$&47.9 &1.2 &47.9 &66.2 &2.3 &67.2 &3.893$\times 10^{33}$ &8.204$\times 10^{33}$   \\
2.4$\times 10^{14}$&54.1 &1.5 &54.1 &72.5 &2.8 &74.0 &5.278$\times 10^{33}$ &1.031$\times 10^{34}$  \\
2.8$\times 10^{14}$&60.0 &1.9 &60.0 &78.3 &3.3 &80.3 &6.819$\times 10^{33}$ &1.251$\times 10^{34}$   \\
3.0$\times 10^{14}$&62.8 &2.1 &62.8 &81.0 &3.5 &83.2 &7.659$\times 10^{33}$ &1.361$\times 10^{34}$   \\
3.4$\times 10^{14}$&68.3 &2.5 &68.3 &86.3 &4.0 &88.9 &9.441$\times 10^{33}$ &1.595$\times 10^{34}$   \\
3.8$\times 10^{14}$&73.5 &2.9 &73.5 &91.2 &4.4 &94.4 &1.136$\times 10^{34}$ &1.835$\times 10^{34}$  \\
4.2$\times 10^{14}$&78.6 &3.3 &78.6 &95.8 &4.9 &99.5 &1.344$\times 10^{34}$  &2.081$\times 10^{34}$  \\
4.6$\times 10^{14}$&83.5 &3.7 &83.5 &100.4 &5.4 &104.4 &1.564$\times 10^{34}$ &2.333$\times 10^{34}$  \\
5.0$\times 10^{14}$&88.3 &4.1 &88.3 &104.6 &5.8 &109.2 &1.798$\times 10^{34}$ &2.591$\times 10^{34}$  \\
5.4$\times 10^{14}$&93.0 &4.6 &93.0 &108.7 &6.3 &113.7 &2.045$\times 10^{34}$ &2.854$\times 10^{34}$   \\
5.6$\times 10^{14}$&95.2 &4.8 &95.2 &110.7 &6.5 &116.0&2.174$\times 10^{34}$  &2.987$\times 10^{34}$ \\
\botrule
\end{tabular}\label{ta3}}
\end{table*}
The signs of `0' and `1' denote $B^{*}\ll 1$ and $B^{*}= 100 $, respectively.
The values of $P^{0}$ and $P^{1}$ are obtained from Eq.(32) and Eq.(34),
respectively. For simplicity, we ignore the change of $Y_{p}$ (or $Y_{e}$)
caused by magnetic effects when calculating $P^{1}$. However, this \textbf{assumption} can not
affect our main conclusion here that the dynamic pressure $P$ increases with
increasing $B$ and $\rho$.  Seeing from Table 3, for an ideal $npe$ system, the
chemical potentials of fermions, as well as matter pressure, increases with
increasing $B$ and $\rho$. Furthermore, the ratios of $P^{1}/P^{0}$ are in the range
of $\sim 2.4~- 1.4$ corresponding to a density range of ($1.4\times 10^{14}-~5.6\times
10^{14}$)g~cm$^{-3}$, but the magnitudes of $P^{1}$ and $P^{0}$ are identical. Based
on Table 3 we plot two schematic sub-diagrams of QED effects on EOS of this $npe$
gas, as shown in Fig.5.
\begin{figure*}[htb]
\begin{center}
\begin{tabular}{cc}
\scalebox{0.78}{\includegraphics{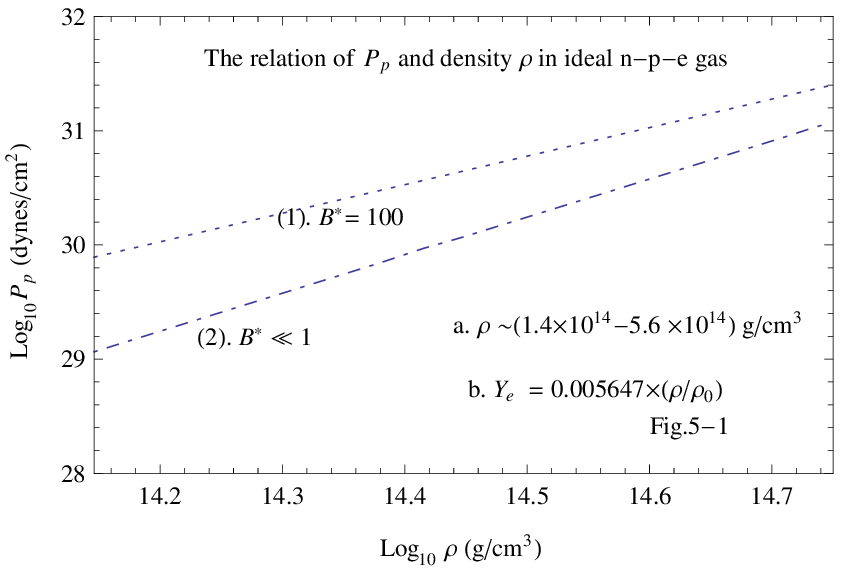}}&\scalebox{0.78}{\includegraphics{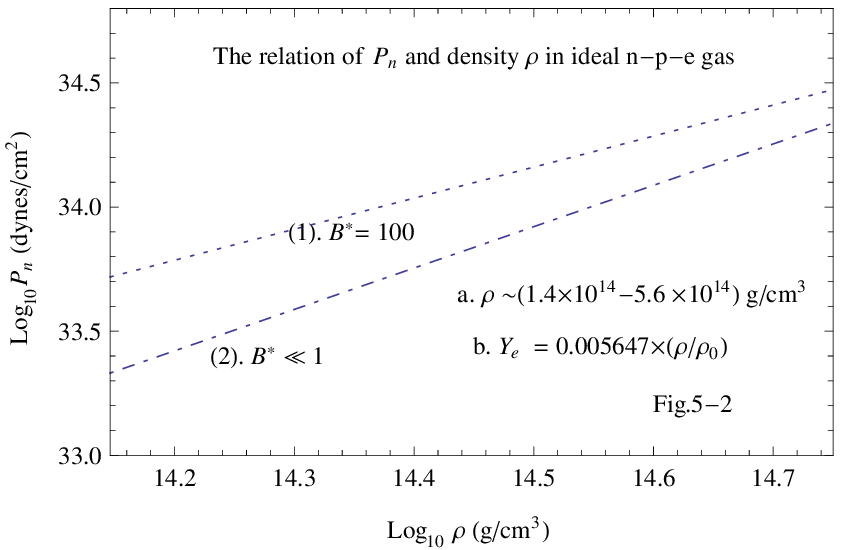}}\\
(a)&(b)\\
\end{tabular}
\end{center}
\caption{The QED effects on $P_{p}$ and $P_{n}$ in ideal $npe$ gas.
Sub-diagrams Fig.5-1 and Fig. 5-2 give $P_{p}$ versus $\rho$ and
$P_{n}$ versus $\rho$, respectively, for an ideal $npe$ gas in a
superhigh magnetic field.  Here the ST approximation is adopted in treating
EOS of this system under $\beta$-equilibrium in the weak-field limit.
\protect\label{fig5}}
\end{figure*}

From sub-diagrams Fig.5-1 and Fig.5-2, both $P_{p}$ and $P_{n}$ increase
obviously with density $\rho$ and magnetic filed $B$. As in magnetic BPS and BBP
models above, given a same density $\rho$, the values of $P_{p}$ and $P_{n}$
in a superhigh magnetic field (e.g.,$B^{*}~= 100 $) are  higher than those in the
weak-field limit.

As we know, in an ideal $npe$ system under strong magnetic fields, there is
a positive correlation between the total matter energy density, $\epsilon$, and the
total matter pressure, $P$,
\begin{equation}
P= -\frac{\partial\epsilon/n}{\partial \frac{1}{n}}= n^{2}\frac{\partial \epsilon/n}{\partial n}~,~~~\label{35}
 \end{equation}
where $n$ is equivalent to the number density of baryons, $P$ includes the
electron pressure, $P_{e}$, proton pressure, $P_{p}$ and neutron pressure,
$P_{n}$. Combining Eq.(34) with Eq.(35), we can conclude that the total matter
energy density, $\epsilon$, increases with increasing $B$.

The stable configurations of a NS can be obtained from the well-known hydrostatic
equilibrium equations of Tolman, Oppenheimer and Volkov (TOV) for the pressure $P(r)$
and the enclosed mass $m(r)$,
\begin{eqnarray}
 &&\frac{dP(r)}{dr}=-\frac{G(m(r)+ 4\pi r^{3}P(r)/c^{2})(\rho+P(r)/c^{2})}{r(r-2Gm(r)/c^{2})}\nonumber\\
&&\frac{dm(r)}{dr} = 4\pi \rho r^{2}~,~~\label{36}
\end{eqnarray}
where $G$ is the gravitational constant, see.Ref.~\refcite{Shapiro83}. For
a chosen central value of $\rho$, the numerical integration of Eq.(23)
provides the mass-radius relation. However, we focus on a qualitative
analysis of relation of $m(r)$ and $B$ in the context of a magnetar. In
Eq.(36), the pressure $P(r)$ is the gravitational collapse pressure, and
always be balanced by the total matter pressure, $P$; the central density
$\rho$ is proportional to the matter energy density $\epsilon$; the
enclosed mass, $m(r)$, increases with the central density $\rho$ when
$r$ is given. From the calculations and analysis in this Section, we
may expect a more massive stelar mass of a magnetars because the matter
energy density, $\epsilon$, increases with $B$. Also, we propose that
magnetars' instability could be associated with the increased gravitational
collapse pressure and high chemical potentials of fermions.

As we know, the magnetic effects can give rise to an anisotropy of
the total pressure of the system to become anisotropic, e.g., see
Ref.~\refcite{Bonazzola93},~\refcite{Bocquet95},~\refcite{Khalilov02},
~\refcite{Perez03},~\refcite{Perez08},~\refcite{Paulucci11}. In an
ideal $npe$ system, the total energy momentum tensor due to both matter
and magnetic field is to be given by
\begin{equation}
T^{\mu \nu}=  T^{\mu \nu}_{m}+ T^{\mu \nu}_{B},~,~~\label{37}
\end{equation}
where,
\begin{equation}
T^{\mu \nu}_{m}= \epsilon_{m}u^{\mu}u^{\nu}- P_{m}(g^{\mu \nu}- u^{\mu}u^{\nu})~,~~\label{38}
\end{equation}
and
\begin{equation}
T^{\mu \nu}_{B}= \frac{B^{2}}{4\pi}(u^{\mu}u^{\nu}- \frac{1}{2}g^{\mu \nu})- \frac{B^{\mu}B^{\nu}}{4\pi}~.~~\label{39}
\end{equation}
where $\epsilon_{m}$ is the total matter energy density, and $P_{m}$ is the total
matter pressure of the $npe$ system, respectively. The first term in Eq.(39) is
equivalent to magnetic pressure, while the second
term causes the magnetic tension. Assuming a uniform magnetic
field $B$ directed along the $z$-axis, then we have
\begin{equation}
T^{\mu \nu}_{B} =
\begin{bmatrix}
\frac{B^{2}}{8\pi} & 0 & 0 & 0 \\
0 & \frac{B^{2}}{8\pi} & 0 & 0 \\
0 & 0 & \frac{B^{2}}{8\pi} & 0 \\
0 & 0 & 0 & -\frac{B^{2}}{8\pi}
\end{bmatrix}.~~\label{40}
\end{equation}
Due to an excess negative pressure or tension along the direction to the
magnetic field, the component of $T^{\mu \nu}_{B}$ along the field, $T^{zz}
_{B}$, is negative. Thus, the total pressure in the parallel direction to
the magnetic field can be written as
\begin{equation}
P_{\|}= P_{m}- \frac{B^{2}}{8\pi},~~\label{41}
\end{equation}
and that perpendicular to the magnetic
field, $P_{\bot}$, is written as
\begin{equation}
P_{\bot}= P_{m}+ \frac{B^{2}}{8\pi}~.~\label{42}
\end{equation}
From Eqs.(41-42), it's obvious that the total pressure of the system becomes anisotropic.
The parallel pressure $P_{\|}$ becomes negative, assuming that the magnetic pressure
exceeds $P_{m}$. However, according to our calculations, when $B^{*}= 100 $, $P_{m}\sim
10^{33}-10^{34}$~dynes~cm$^{-2}$ and $\frac{B^{2}}{8\pi}\sim 10^{29}-10^{30}$. Hence, in
this presentation, we consider that the component of the total energy
momentum tensor along the symmetry axis becomes positive, $T^{zz}> 0$, since the total
matter pressure increases more rapidly than the magnetic pressure. Accordingly, the total
energy momentum tensor is given by
\begin{equation}
T^{\mu \nu} =
\begin{bmatrix}
\epsilon_{m}+\frac{B^{2}}{8\pi}& 0&0& 0 \\
0&P_{m}+\frac{B^{2}}{8\pi}&0&0 \\
0&0&P_{m}+\frac{B^{2}}{8\pi}& 0 \\
0&0&0&(P_{m}+\frac{B^{2}}{8\pi})-\frac{B^{2}}{4\pi}
\end{bmatrix}.~\label{43}
\end{equation}
A strong magnetic field can uncover anisotropy, see Ref.~\refcite{Perez03},
~\refcite{Perez08}, due to magnetization pressure, ${\cal M}B$, where ${\cal M}$
is the magnetization of the system, which is given by
\begin{equation}
{\cal M}= - \frac{\partial \epsilon_{m}}{\partial \textit B}.~\label{44}
\end{equation}
Therefore, the pressure perpendicular to the magnetic field, $P_{\bot}$,
is actually written as
\begin{equation}
P_{\bot}= P_{m}+ \frac{B^{2}}{8\pi}- \cal M \textit B.~\label{45}
\end{equation}

For a magnetic field $B\ll 10^{20}$ G, the magnetization is opposite to
the external field $B$, i.e., ${\cal M}< 0$, and it may happen that
$P_{\bot }> P_{\|}$.  For a magnetic field $B\geq 10^{20}$ G, the
opposite occurs in some permeable materials where ${\cal M}> 0$ and
$P_{\bot }< P_{\|}$, this is due to ferromagnetic effects which have
quantum origin, as in a gas of degenerate neutrons (see Ref.
~\refcite{Perez03},~\refcite{Perez08}).

However, magnetars considered in the present work universally have
typical surface dipole magnetic fields $\sim(10^{14}-10^{15})$ G and
inner field strengths not more than $10^{17}$ G, under which
$B^2/8\pi \gg \cal M \textit B$, and the effects of AMMs of
nucleons on the EOS are ignored. Therefore, we exclude
magnetization term in the total pressure. In other words, the
exclusion of magnetization term cannot affect the result practically
for the present purpose. Our model of magnetized ideal $npe$ gas
favors the following relation: $P_{\bot}> P_{\|}$, which could lead
to the Earth-like oblatening effect.

In the work of Bocquet et al.(see Ref~\refcite{Bocquet95}), the authors
considered an extension of the electromagnetic code developed by Bonazzola et al.
(see Ref.~\refcite{Bonazzola93}), and simulated high-magnetized rotating NSs.
According to their simulations, the component of the total energy momentum
tensor along the symmetry axis is negative, and the combined fluid-magnetic
medium can develop a magnetic tension. As a result of this tension, the star
displays a pinch across the symmetry axis and assumes a flattened shape (see
Ref.~\refcite{Bocquet95}).

Contrary to the work of Bocquet et al.(1995), we propose that the component
of the total energy momentum tensor along the symmetry axis becomes positive,
since the total matter pressure always grows more rapidly than the magnetic
pressure. However, a similar effect is expected to occur in our present work,
where the magnetic tension along the direction to the magnetic field will be
responsible for deforming a magnetar along the magnetic field, and turns the
star into a kind of oblate spheroid. Be note that such a deformation in shape
might even render a more compact magnetar endowed with canonical strong surface
fields $B\sim 10^{14-15}$~G. Also, such a deformed magnetar could have a more
massive mass because of the positive contribution of the magnetic field energy
to the EOS of the system.

\section{Two contrary views on the relationship between $P_{e}$ and $B$}
As we know, the electron pressure,$P_{e}$, as well as the electron
Fermi energy $E_{\rm F}(e)$, is generally believed to decrease with
increasing magnetic field strength $B$ (e.g., Ref.~\refcite{Lai91},~\refcite{Lai01}).

According to statistical physics, the microscopic state number in a 6-dimension
phase-space element $dxdydzdp_{x}dp_{y}dp_{z}$ is $dxdydzdp_{x}dp_{y}dp_{z}/h^{3}$.
In the Ref.~\refcite{Canuto68}, the number of states occupied by completely
degenerate relativistic electrons in an unit volume is calculated as follows:
\begin{eqnarray}
 &&N_{phase}= \sum_{p_{x}}\sum_{p_{y}}\sum_{p_{z}}
 = \frac{1}{h^{3}}\int_{-\infty}^{\infty}\int_{-\infty}^{\infty}\int_{-\infty}^{\infty}dp_{x}dp_{y}dp_{z}\nonumber\\
 &&= \frac{1}{h^{3}}\int_{0}^{p_{\rm F}}dp_{z}\int_{0}^{\infty}p_{\bot}dp_{\bot}\int_{0}^{2\pi}d\Phi
 = \frac{\pi p_{\rm F}}{h^{3}}\int_{0}^{\infty}dp_{\bot}^{2}~~,\label{46}
 \end{eqnarray}
where $\Phi~=~tan^{-1}p_{y}/p_{x}$. Quantization requires $p_{\bot}^{2}\rightarrow m^{2}c^{4}\frac{B}{B_{\rm cr}}2n$,
hence, $\int_{0}^{\infty}dp_{\bot}^{2}\rightarrow \sum_{n=0}^{\infty}\omega_{n}$,
where $\omega_{n}$ is the degeneracy of the $n$-th Landau level of electrons in relativistic
magnetic field, which can be estimated as
\begin{eqnarray}
&&\omega_{n}= \frac{1}{h^{2}}\int_{0}^{2\pi}d\phi\int_{A < p_{\bot}^{2}< B}p_{\bot}dp_{\bot}\nonumber\\
&&= \frac{2\pi}{h^{2}}\frac{(B-A)}{2}= \frac{1}{2\pi}(\frac{\hbar}{m_{e}c})^{-2}\frac{B}{B_{\rm cr}}~~,\label{47}
\end{eqnarray}
where $A= m^{2}c^{2}\frac{B}{B_{\rm cr}}2n$, and $B= m^{2}c^{2}
\frac{B}{B_{\rm cr}}2(n+1)$ (see Ref.~\refcite{Canuto71})

Be note, the above calculation method was cited from \textit{Statistical
Mechanics (1965)}(see Ref.~\refcite{Kubo65}). The classical textbook,
\textit{Statistical Mechanics (2003)}(see Ref.~\refcite{Pathria03}), also
gives the expression for the degeneracy of the $n$-th Landau level of
electrons in a relativistic magnetic field,
\begin{eqnarray}
 &&\omega_{n}= \frac{1}{h^2}\int dp_{x}dp_{y}= \frac{1}{h^2}\pi p_{\bot}^{2}\mid_{n}^{n+1}\nonumber\\
&&= \frac{4\pi m_{e}\mu_{e}B}{h^{2}}=\frac{1}{2\pi}(\frac{\hbar}{m_{e}c})^{-2}\frac{B}{B_{\rm cr}}~~,\label{48}
 \end{eqnarray}
where $\mu_{e}^{'}=\frac{e\hbar}{2m_{e}c}$ is the electron magnetic
moment. In the momentum interval~$\Delta p_{z}$~along the direction to the magnetic field,
for a non-relativistic electron gas, the number of possible microstates is given by
\begin{equation}
 \frac{{e}BS}{2\pi\hbar c}\frac{L_{z}}{2\pi \hbar}\Delta p_{z}= \frac{{\rm e}BV}{4\pi^{2}\hbar^{2}c}~~.~~\label{49}
 \end{equation}
(see Ref.~\refcite{Landau65}). For the convenience of calculation, we assume $S=~1$~cm$^{2}$ and~$V=~1$~cm$^{3}$,
and consider the electron spin degeneracy $g_{n0}= 2- \delta_{n0}$~(when $n = 0$, $g_{00}= 1$; when~$n \geq 1$, $g_{n0}= 2$).
From \textbf{Eq.(49)}, we get the expression for the degeneracy of the n-th Landau level of electrons in a non-relativistic magnetic field,
\begin{equation}
 \omega_{n}= g_{n0}\frac{eB}{2\pi \hbar c}= g_{n0}\frac{4\pi m_{e}\mu_{e}B}{h^{2}}~~,~\label{50}
 \end{equation}
where the solution of the non-relativistic electron cyclotron motion,
$\hbar\omega_{B}= 2\mu_{\rm e}B$, is used. Confusingly, the expression
for the degeneracy of the $n$-th Landau level of electrons in a relativistic
magnetic field is completely in accordance with that for the corresponding
non-relativistic case, if one compare Eq.(47)(or Eq.(48)) with Eq.(50).
After careful consideration and analysis, we find that Eq.(47) (or Eq.(48))
is just the expression we are looking for, that leads to an incorrect
viewpoint on the electron Fermi energy, as well as the electron pressure,
prevailing in the world currently. In our opinion, the expression for
$\omega_{n}$ in Eq.(47)(or Eq.(48)) is incorrect, because it's against the
viewpoint on the quantization of Landau levels. The essence of the above
method (or the incorrect deduction) lies in the assumption that the torus
located between the $n$-th Landau level and the $(n+1)$-th Landau level in
momentum space is ascribed to the $(n+1)$-th Landau level. Such a factitious
assumption is equivalent to allow a continuous momentum (or energy) of an
electron moving in the direction perpendicular to the magnetic field, which
is obviously contradictory to the quantization of Landau levels in the case
of a strong magnetic fields. The concept of Landau level quantization clearly
tells us that there is no any microscopic quantum state between $p_{\bot}(n)$
and$p_{\bot}(n+1)$. In a word, the main cause of the popular incorrect
viewpoint on $E_{\rm F}(e)$ is due to a factitious assumption.

In order to depict the quantization of Landau levels truly and accurately, we must
introduce the Dirac $\delta$-function $\delta(\frac{p_{\bot}}{m_{e}c}-~[2(n+\sigma
+ \frac{1}{2})B^{*}]^{\frac{1}{2}})$~(see Ref.~\refcite{Gao11a},~\refcite{Gao12a}).
As we know, the eigenvector wave function of the Schr\"{o}dinger equation (or Dirac
equation) can be expanded in an infinite series. In the process of deducing the
expressions concerning the quantization of Landau levels, we should firstly give a
$p_{z}$ that changes continuously along the direction to the magnetic field, then
solve the relativistic Schr\"{o}dinger equation (or Dirac equation), finally obtain
the maximum  Landau level number, $n_{max}$, by truncating the infinite series when
the wave function is limited (see Ref.~\refcite{Landau65}). Logically, give $p_{z}$
firstly, and then determine the maximum Landau level number $n_{max}$.

As an alternative way to depict the quantization of Landau Levels of electrons in strong
magnetic fields, we rewrite Eq.(50) as
\begin{eqnarray}
 &&\omega_{n}= \frac{1}{h^{2}}g_{n0}\int_{0}^{2\pi}d \phi \int \delta(\frac{p_{\bot}}{m_{e}c}-[2(n+ \sigma+ \frac{1}{2})B^{*}]^{\frac{1}{2}})
 p_{\bot}dp_{\bot} \nonumber\\
 &&= \frac{2\pi}{h^{2}}g_{n0} \int \delta(\frac{p_{\bot}}{m_{e}c}-~[2(n+ \sigma+ \frac{1}{2})B^{*}]^{\frac{1}{2}})p_{\bot}dp_{\bot}~~,~\label{51}
 \end{eqnarray}
where $\phi= \arctan p_{y}/p_{x}$, and $g_{n0}= 2- \delta_{n0}$ (when $n = 0$,
$g_{00}= 1$; when~$n \geq 1$, $g_{n0}= 2$)is the electron spin degeneracy.
In the interior of a NS, in the light of the Pauli exclusion principle,
the electron number density should be equal to its microscopic state density,
\begin{eqnarray}
&&N_{phase}= n_{e}= 2\pi \frac{(m_{e}c)^3}{h^{3}}\int_{0}^{\frac{E_{\rm F}(e)}{m_{e}c^2}}d(\frac{p_{z}}{m_{e}c})\sum_{n=0}^{n_{max}
(p_z, \sigma, B^{*})}\sum g_{n0}\nonumber\\
 &&\int_{0}^{\frac{E_{\rm F}(e)}{m_{e}c^2}}\delta(\frac{p_{\bot}}{m_{e}c}
 -[2(n+ \sigma + \frac{1}{2})B^{*}]^{\frac{1}{2}})\frac{p_{\bot}}{m_{e}c}d(\frac{p_{\bot}}{m_{e}c})= N_{A}\rho Y_{e}~~.~\label{52}
 \end{eqnarray}
In a word, when calculating $\omega_{n}$ and $N_{phase}$, we must take into
account the Dirac $\delta$-function, otherwise, we would reach the wrong
conclusion that $E_{\rm F}(e)$ decreases with the increase in $B$. Be note,
in this letter, we first time propose that the electron pressure $P_e$
increases with increasing $B$, and the popular point of view on $P_e$ will
be confronted with a severe challenge from our calculations

\section{Summary Conclusions}
In this paper we have derived a general expression for pressure
of degenerate and relativistic electrons, which holds approximately when
$B\gg B_{cr}$. We conclude that the stronger the magnetic field is, the higher
the electron pressure becomes. The high electron pressure could be caused by
the electron Fermi energy, which increases with the increase in the magnetic
field strength. Given these arguments, the popular point of view on $P_{e}$ will
be confronted with a severe challenge from our calculations.

Also, we have taken into account QED effects on EOSs of $BPS$ model, $BBP$ model
and ideal $npe$ gas model, and have discussed an anisotropy of the total pressure of
ideal $npe$ gas due to strong magnetic fields. Magnetars we have adopted in this
work have typical surface dipole fields $B_{dip}\sim(10^{14}-10^{15})$ G, under
which the effects of AMMs of nucleons on the total energy density and the total
pressure have not been considered. The main conclusions are as follows: The total
matter pressure $P$ increases with magnetic field strength $B$, because chemical
potentials of fermions increase with increasing $B$, given a matter
density $\rho$; taking into account of QED effects on the EOS, the total pressure
is anisotropic; comparing with a common radio pulsar, a magnetar could be a more
compact oblate spheroid-like deformed NS, due to the anisotropic total pressure;
an increase in the maximum mass of a magnetar is expected because of the positive
contribution of the magnetic field energy to the EOS.

Finally, it is earnestly hoped that our calculations can soon be combined with
the latest studies and observations of magnetars, to present a deeper understanding
of the nature of superhigh magnetic fields and bursts of magnetars.

\section*{Acknowledgments}
We thank the anonymous referee for the care in reading the manuscript and
for valuable suggestions which help us to improve this paper substantially.
This work is supported by Xinjiang Natural Science Foundation No.2013211A053.
This work is also supported in part by Chinese National Science Foundation through grants
No.11173041, No.11003034 and No.11133001, National Basic Research Program of China grants 973 Programs 2009CB824800 and 2012CB821800, and by  a research fund from the Qinglan project of Jiangsu Province.

\end{document}